\documentclass[12pt]{article}
\usepackage								{textcomp}                         
\usepackage[utf8]                       {inputenc}
\DeclareUnicodeCharacter{00A0}			{~}
\usepackage[T1]                         {fontenc}
\usepackage								{tgbonum}
\usepackage								{geometry}
\usepackage								{charter}
\usepackage								{MnSymbol}
\DeclareUnicodeCharacter				{00A0}{ }
\usepackage[toc,page]					{appendix}
\usepackage                             {color}
\usepackage                             {amsmath}
\usepackage                             {graphicx}
\usepackage			       				{booktabs}
\usepackage[flushleft]					{threeparttable}
\usepackage[english]                    {babel}
\usepackage                             {csquotes}
\usepackage                             {mathrsfs,amsmath}
\usepackage                             {relsize}
\usepackage[nottoc]                     {tocbibind}
\usepackage                             {subcaption}
\usepackage								{soul}
\usepackage                             {physics}
\usepackage								{float}
\usepackage								{units}
\DeclareUnicodeCharacter{2010}{-}% support older LaTeX versions
\usepackage								{rotating}
\usepackage								{tikz}
\usepackage								{adjustbox}
\usepackage								{blindtext}
\usepackage                             {hyperref}
\definecolor							{darkblue}{rgb}{0.0,0.0,0.4}
\definecolor							{darkgreen}{rgb}{0.0,0.4,0.0}
\hypersetup{
    colorlinks=true,
    linkcolor=blue,
    menucolor=red,
    citecolor=darkgreen,
    urlcolor=darkgreen,
    filecolor=magenta,
}
\usepackage{url}
\usepackage{array}
\usepackage{makecell}
\usepackage{sidecap}
\sidecaptionvpos{figure}{c}
\usepackage{tcolorbox}
\usepackage{titling} % Customizing the title section
\usepackage{times}
\usepackage{lineno}

\title{{Sub-femtosecond optical control of entangled states}\\ %- \\ \textcolor{red}{A consequence of molecular entanglement}
} 

\author
{Farshad Shobeiry$^{ \ast,1}$, 
Patrick Fross$^{1}$, 
Hemkumar Srinivas$^{1}$,\\  
Thomas Pfeifer$^{ \ast,1}$, 
Robert Moshammer$^{ \ast,1}$,   
Anne Harth$^{ \ast,1,2}$  \\
\\
\normalsize{1 Max Planck Institute for Nuclear Physics, Saupfercheckweg 1, 69117 Heidelberg}\\
\normalsize{2 Center for Optical Technologies, Aalen University, Anton Huber Straße 1, 73430 Aalen}\\
\\
\\
\\ \normalsize{$^\ast$Corresponding authors. Email: shobeiry@mpi-hd.mpg.de, anne.harth@hs-aalen.de,}
\\ \normalsize{ thomas.pfeifer@mpi-hd.mpg.de, r.moshammer@mpi-hd.mpg.de. }
}

\date{}

\begin{document} 
\baselineskip24pt
\maketitle 
\newpage

\noindent{\large \textbf{Abstract}}
\noindent 
For photo-dissociation of a single hydrogen molecule ($H_2$) with combined XUV and IR laser pulses, we demonstrate optical control of the emission direction of the photoelectron with respect to the outgoing neutral fragment (the H-atom). 
Depending on the relative delay between the two laser fields, adjustable with sub-femtosecond time resolution, the photoelectron is emitted into the same hemisphere as the H-atom or opposite. 
This emission asymmetry is a result of entanglement of the two-electron final-state involving the spatially-separated bound and emitted electron.  

\vspace{10pt}
\noindent{\large \textbf{Main}}
Attosecond physics has developed into a central field that enables research into ultrafast coherent electron dynamics in matter, e.g. triggered by photoexcitation and photoionisation. 
A fundamental observation in this field is the emission behavior of photoelectrons after the dissociation of molecular hydrogen ($H_2$). 
Normally, the emission direction of a photoelectron is symmetric relative to the ejected neutral hydrogen atom, with no preferred direction, if the inversion symmetry (parity) of the electron wave function is maintained during the process. 
However, when the molecule is set in a superposition of states with opposite parities, an asymmetric electron emission pattern emerges \cite{Bello2021}. 
This phenomenon emphasizes the interplay between molecular structure and electron dynamics. 
In order to thoroughly investigate the coherent dynamics triggered  by photoionisation, in particular using the combination of attosecond laser pulses in the extreme ultraviolet (XUV) and infrared (IR) pulses, a detailed analysis of the resulting entanglement of ions and photoelectrons is essential and provides insight into the fundamental processes of attosecond physics.

Theoretical work is in progress that investigates the ultrafast coherent control of entangled states, e.g. \cite{Vatasescu2023,Ruberti2019,Nishi2019, Carlstrom2018,Nabekawa2023,Vrakking2021}. 
Kroll et al. \cite{Kroll2022} discussed experiments on controlling quantum entanglement between an ion and a photoelectron in attosecond pump-probe experiments measuring the ion kinetic energy via a velocity imaging mapping device, by tailoring the spectral properties of attosecond XUV laser pulses.  
The research demonstrates how the degree of entanglement - and consequently, the vibrational coherence of the ion - can be manipulated.
In our experiment, we demonstrate the asymmetric emission in the bipartite system of photoelectron/ion by detecting them in coincidence while they are spatially separated and steer their {relative emission}  direction using the delay between XUV and weak IR pulses.

This effect should not be confused with the localization of the bound electron on one of the nuclei in the \textit{laboratory frame} using a spatially asymmetric strong laser field during dissociation as reported in \cite{PhysRevLett.103.213003,Sansone2010,Kling246,PhysRevLett.104.023001}. 
In those experiments, only the proton (deuteron) is detected and its ejection direction with respect to the laser polarization is controlled by the phase of the light field irrespective of the direction of the ejected photoelectron. 
%Sansone’s experiment in 2010 showed asymmetry in ion emission related to XUV-IR pulse delays, measuring only protons. 
Our study advances this by measuring both photo-electrons and protons in \textit{coincidence}, assessing the emission direction in the \textit{molecular frame}. 
%Previews experiments that measure an emission asymmetry as a result of entanglement of the two-electrons, including those by Sansone \cite{Sansone2010} and Kling \cite{Kling246}, did not claim to measure entanglement directly. 
Moreover, in contrast e.g. to Sansone's experiment, where the lab-frame asymmetry was analyzed under the condition that ions appear with large kinetic energies  - this is indicative of a fragmentation mechanism that involves the \textit{decay of doubly-excited states} in the molecule - ,  our study focuses on \textit{ground-state fragmentation} via coherent superpositions of mixed-parity states. 
This approach is applicable to experiments using weak IR fields that do not introduce lab-frame asymmetry, and we observed no lab-frame asymmetry for either photoelectrons or ions. 
Finally, in certain photoninduced molecular fragmentation scenarios, a preferential photoelectron emission direction in the molecular frame suggests entanglement \cite{Fischer2013}, though such identification is complex and M. Vrakking emphasizes that true entanglement detection necessitates coincident two-particle measurement, moving beyond traditional attosecond experiment protocols that measure photo-electrons or ions separately \cite{Vrakking2022}.
%: In such experiments “… a well-defined ion + photoelectron wavefunction is formed, which can be calculated by solving the TDSE, or about which extensive information can be obtained using a fragment ion + photoelectron coincidence measurement. Such an elaborate experimental protocol significantly goes beyond the experimental protocols that are typically used in attosecond pump-probe experiments, where experiments are commonly limited to measurements of either the photoelectrons or (fragment) ions that are formed in the experiment.” [Vrakking2022].
This asymmetric photoelectron emission direction is sensitive to a phase difference between different dissociative pathways coming from e.g. opposite parity states.
So far, this effect has been observed by single-photon dissociative photo-ionization \cite{Martin629,Fischer2013}. 

Our work overcomes experimental challenges to \textit{control entanglement} by manipulating the relative phases between the different pathways from outside, achieving control over molecular frame asymmetry at a sub-femtosecond scale. 
We demonstrate active steering of the asymmetric electron emission by using a \textit{few-photon} interaction with different colors in combination with the control of their relative delay. 
This capability is the core achievement of our study.

%The emission direction of a photoelectron after dissociation of molecular hydrogen $H_2$ is symmetric and shows no preferred direction with respect to the ejected neutral hydrogen atom if the inversion symmetry (parity) of the electron wave function remains during the process. However, the molecule can be placed in a superposition of states with opposite parities which leads to an asymmetric electron emission \cite{Bello2021}. \textcolor{red}{This asymmetric emission is sensitive to a phase difference between different dissociative pathways coming from e.g. opposite parity states.} So far, this effect has been observed by \textit{single-photon} dissociative photoionization \cite{Martin629,Fischer2013}.However, in these experiments it is not possible to actively control the phase shift between the different pathways from outside.In this manuscript, we demonstrate active steering of the asymmetric electron emission by using a \textit{few-photon} interaction with different colors in combination with the control of their relative delay. 

\vspace{20pt}
%\textcolor{gray}{Experimental set-up -- gray text will be removed later}\\
We determine the emission asymmetry from the angular distribution of the photoelectron with respect to the ejected neutral hydrogen atom containing the bound electron after the $n$-photon dissociation of $H_2$: 
\begin{equation}
H_2+n\gamma\;\rightarrow \; H^+\;+\;H \;+\;e^-\label{disso}
\end{equation}
by detecting the electron $e^-$ and proton $H^+$ in coincidence using a reaction microscope (REMI) apparatus \cite{remi}, where the neutral hydrogen atom $H$ is ejected opposite to the direction of the proton. $\gamma$ stands for the respective photon energies and $n$ is the number of interacting photons. 
The  spectrometer consists of two position-sensitive detectors for electrons and ions. 
We can retrieve the 3D momentum components of the ion $H^+$ and the electron $e^-$ using time-of-flights and hit positions on the detectors (see supplementary for more details).  

Our light source is a femtosecond laser with a central wavelength of \unit[1030]{nm} (IR) and a pulse duration of \unit[50]{fs}. 
The beam is split into two arms. 
The main part is used to produce a comb-like ultraviolet (XUV) {spectrum off odd harmonics with} energies up to \unit[40]{eV} using the process called high-harmonic generation \cite{Paul1689}. 
A very schematic representation of a part of such a comp-like spectrum is visualized in blue and labeled with "$q+2$", "$q$" and  "$q-2$" in Fig.\ref{fig:h2paths}. 
The equally-separated maxima have an energy difference which is twice the IR photon energy (2 $\times$ \unit[1.2]{eV}). 
The remaining weaker part is delayed in time and overlapped spatially and temporally with the XUV pulses.
{Since any phase jitter between the XUV and IR pulses is critical for the experiential results, the beam line can be actively stabilized \cite{Hem2023}.} 
The basic experimental principles are similar to \cite{Cattaneo2018}.
Both pulses interact  with a gas jet of H$_2$ in the REMI.

\begin{figure*}
\centering 
\includegraphics[keepaspectratio=true,scale=.98]{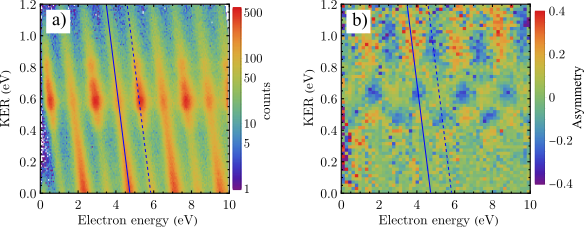}
\caption{\textbf{Interaction of XUV and IR photons with a hydrogen molecule: energy sharing spectrograms and asymmetry histograms.}  \textbf{a)}, Joint energy spectrum  (JES) for the dissociative ionization of H$_2$.  One example of an "odd band" (OB) is indicated with a diagonal solid line. This specific OB is a result of the absorption of XUV photons with an energy of \unit[22.8]{eV}. The absorption of XUV+IR photons results in "even bands" (EB). An exapmle of such an EB is marked with a diagonal dashed line. The enhancement of the signal at a KER of \unit[0.6]{eV} is due to the absorption of an IR photon by the molecular ion (bond softening). \textbf{b)}  The asymmetry parameter in the case of dissociation with XUV and IR photons. The asymmetry is non-zero in the region between KER \unit[0.35]{eV} to \unit[1.2]{eV}. The two diagonal lines, indicating the position of the marked OB and EB from a) and serve as a guide.}
\label{fig:viblevel}
\end{figure*}

\vspace{10pt}
%\textcolor{gray}{Static symmetry -- gray text will be removed later}\\
The main process is single-ionization of H$_2$ into the molecular ion   ground state (1$s\sigma_g$);
only a small fraction  of ionization events lead to dissociation for photons with energies higher than the dissociation limit of H$_2$ (I$_d$ = \unit[18.1]{eV}). 
As described above, with the REMI we are able to obtain the energy and momentum (\textbf{K}) of both - the proton and the electron - for each dissociation event in coincidence.
The energy of the not detected neutral hydrogen atom is reconstructed using momentum conservation: \textbf{K}$_{H}+$\textbf{K}$_{H^+}+$\textbf{K}$_{e^-}=0$. 
After dissociation, unlike atoms, the energy of the incoming photon can be  distributed among electronic and nuclear degrees of freedom:
\begin{equation}
\overbrace{E_{H^+}+E_H}^\text{KER}+E_{e^-}= E_{\gamma}-I_d. \label{energybalance}
\end{equation} 
The sum of the energy of the proton and hydrogen atom ($E_{H^+}+E_H$) is referred to as the kinetic energy release (KER).   
In order to visualize the distribution of the absorbed photon energy between nuclei and the electron, we plot the KER versus the electron energy ($E_{e^-})$ in a 2D histogram (joint energy spectrum (JES)). 
%\textcolor{blue}{For these data, the delay between the XUV and IR pulses where varied too; stimmt das?} 
Fig. \ref{fig:viblevel} a) shows a JES for photodissociation of H$_2$ with a combination of XUV and IR pulses in a KER region from \unit[0]{eV} to \unit[1.2]{eV}.
%for a fixed time delay between the pulses \textcolor{blue}{(oder ist es gemittelt über die Zeit?). }

Due to energy conservation, most events appear on diagonal lines with a slope of -1. 
Two types of such lines exist. We call them "odd bands" (OBs), mainly caused by single-XUV-photon absorption and "even bands" (EBs) caused by XUV-IR two-photon absorption. 
For example, absorption of XUV photons with  an energy of \unit[25.2]{eV} results in a total energy of \unit[22.8]{eV}-\unit[18.1]{eV}= \unit[4.7]{eV} shared between electron and nuclei resulting in events on an OB highlighted by the diagonal solid line (Fig. \ref{fig:viblevel} a)). 
A combination of XUV and IR photons results in an EB marked with a diagonal dashed line.  
The enhancement of the dissociation signal at a KER  of around \unit[0.6]{eV} is due to a process known as bond softening \cite{PhysRevLett.64.1883}.

In order to quantify an asymmetric electron emission, we define an asymmetry parameter $A$: 
\begin{equation}
A=\frac{N_{\theta<90}-N_{\theta>90}}{N_{\theta<90}+N_{\theta>90}},\label{asymmetryequation}
\end{equation}
with $\theta$ being the electron emission angle with respect to the ejected proton as shown in the small box in Fig. \ref{fig:h2paths}. 
$N_{\theta<90}$ and $N_{\theta>90}$ are the numbers of events where the electron and proton are emitted in the same and the opposite hemisphere, respectively. 
A plot of the asymmetry parameter $A$ for the events in Fig. \ref{fig:viblevel}a) is shown in Fig. \ref{fig:viblevel}b). 
In a KER region from \unit[0.4]{eV} to \unit[1]{eV}, we clearly observe that the electron has a preferential direction with respect to the emitted proton ($A$ is non-zero) for all electron energies. 
Additionally, OBs and EBs show different trends!

We like to note that the asymmetry A (see Eq. \ref{asymmetryequation} and Fig. \ref{fig:viblevel}b)\,) becomes visible only in the molecular frame or, in other words, when the relative emission-angle between electron and proton is plotted. When we analyze the proton emission in the lab-frame, without taking the electron data into account, we do not observe any asymmetry at all. This is not surprising because the acting light-field does not exhibit any asymmetry as well. It consists of an IR pulse with many cycles and an XUV field of odd harmonics where the individual XUV pulses appear in sync at each half-cycle of the IR-field. A clear lab-frame asymmetry in the proton emission can be measured \cite{PhysRevLett.104.023001} if XUV pulses containing both odd and even harmonics are used. In this case, the symmetry of the light-field is broken because the XUV pulses appear at every second half-cycle of the IR-field, at instants when the electric field of the IR pulse is pointing into the same direction.

\vspace{10pt}
%\textcolor{gray}{Where does the static asymmetry come from? - Entanglement -- gray text will be removed later}\\
{The energy-dependent asymmetry in \ref{fig:viblevel}b) is most pronounced in the region}  where two dissociation contributions overlap: the ground-state dissociation (contribution 1) and the bond softening (contribution 2).
With the presence of many distinct XUV photon energies, several quantum paths within the two contributions exist and will interfere. 
However, only one pathway for each contribution is represented exemplary in Fig. \ref{fig:h2paths}. 
\begin{figure*}
\centering 
\includegraphics[keepaspectratio=true,scale=1.05]{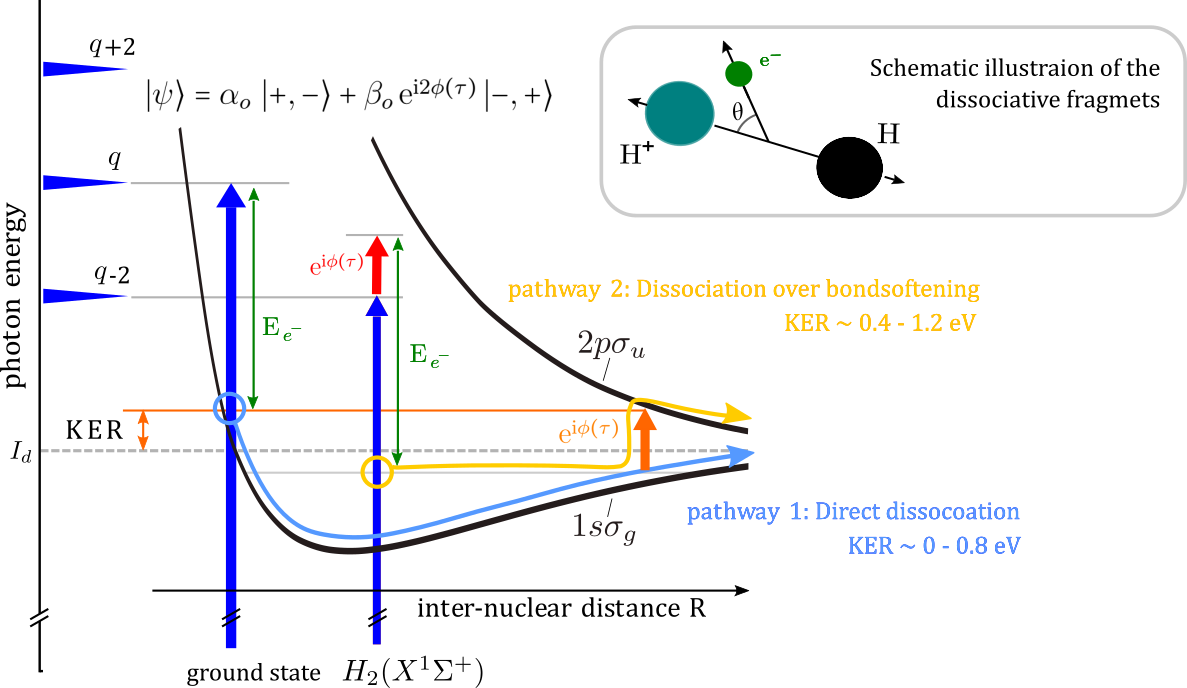}
\caption{Schematic illustration of two dissociation channels of H$_2$ with a combination of XUV (blue arrows) and IR (red and orange arrows) photons. The ground state of H$_2$ ($X^1\Sigma^+_g$)  is a well defined state denoted with respect to its parity $\ket{+,+}$.  The ground state ($1s\sigma_g$) and the first excited state ($2p\sigma_u$) of the molecular ion (H$_2^+$) are shown with black curves. The dissociation limit ($I_d$) is shown with the horizontal gray dashed line. Two indistinguishable dissociation pathways 1 (light blue curve, belonging to ground-state dissociation) and 2 (yellow curve belonging to bond softening) add up coherently to the same final state, namely the same photoelectron energy (E$_{e^-}$, green arrows) and KER. The inset shows the schematic illustration of the dissociative fragments: electron, proton and neutral hydrogen atom.$\theta$ is the emission angle of the photoelectron with respect to the H$^+$ emission direction.}
\label{fig:h2paths}
\end{figure*}
In case of OBs (see supplementary material for EBs),
in pathway 1 (light blue curve), an XUV photon with an energy $\gamma_q=q\,\hbar \omega$ (where $\omega=$\unit[1.2]{eV} is the photon energy of the laser radiation and $q$ labels a maximum in the XUV spectrum) higher than $I_d$  can lead to dissociation along the  1s$\sigma_g$ state resulting in KERs < \unit[2]{eV}. 
This pathway leads to an electron energy of $E_{e^-}=\gamma_q-I_d - $ KER; the parity of the bound and the continuum electron becomes  gerade $\ket{+}$ and ungerade $\ket{-}$, respectively.

In pathway 2 (yellow curve), the molecule is first ionized with the next lower harmonic photon ($\gamma_{q-2}= (q-2)\,\hbar\omega$), to a vibrational level of the ground state ($1s \sigma_g $) of the molecular ion. 
This process happens in the presence of a weak IR probe field where the photoelectron absorbs an IR photon instantaneously leading to a photoelectron energy of $E_{e^-}=\gamma_{q-2}+\hbar \omega -E_b$, where $E_b$ is the energy of the bound vibrational level. 
The symmetry of the emitted photoelectron after absorption of one IR photon becomes gerade $\ket{+}$. 
Further, the molecular ion, containing the bound electron, absorbs at a higher inter-nuclear distance R an additional IR photon leading the molecule to dissociate along the repulsive $2p\sigma_u$ ionic state.
Thus, the bound electron is left with an ungerade symmetry $\ket{-}$.
     
The condition for quantum interference is fulfilled, when the total energy of both, the electron and the KER, are the same for both contributions. 
Hence, the final wave function can be written as a coherent superposition
\begin{equation}
 \ket{\psi}= \alpha_o \ket{+,-}+\beta_o \rm{e}^{i2\phi(\tau) }, \ket{-,+}
 \label{eq:bellv2}
\end{equation}
where $\alpha_o$ and $\beta_o$ are complex probability amplitudes. 
Here we use the notation:
\begin{center}$\ket{\textsc{parity} \text{ of the } \textsc{bound} \text{ electron}, \textsc{parity} \text{ of the } \textsc{free} \text{ electron}}$.\end{center}
$\phi(\tau)$ is an XUV-IR pulse time-delay dependent phase shift.
{The vibrational nuclear wave-function can be omitted in the final wave function, since it is the same for both contributions for the same KER and always of gerade symmetry in case of H$_2$.}
{Eq. 4 resembles the well-known Bell states: $\ket{\psi_B^{\pm} }= \frac{1}{\sqrt{2}}( \ket{+,-} \pm \ket{-,+})$.
And indeed, the observation of an asymmetric emission direction of the photoelectron with respect to its ion indicates that the two electrons - the photoelectron and the still bound electron - originating from the $H_2$ molecule are entangled with respect to parity \cite{Fischer2013}.
}

\vspace{10pt}
%\textcolor{gray}{time dependent asymmetry -- gray text will be removed later}\\
{Since our experimental design allows to vary the phase $\phi(\tau)$ by changing the time-delay between the XUV and IR pulses, we are able to investigate a delay dependent asymmetry parameter $A$.}
In order to show the time-dependence, we add all the OBs together and subtract the time-averaged asymmetry (see supplementary material). 
The resulting asymmetry is plotted as a function of time-delay in Fig. \ref{fig:asymtimedep}a). 
The oscillation of the asymmetry parameter $A$ as a function of the delay with a periodicity of \unit[1.7]{fs} at a given KER demonstrates the sub-femtosecond laser control of the phase $\phi$ of the entangled states of Eq. \ref{eq:bellv2}. 

\begin{figure}
\centering 
\includegraphics[keepaspectratio=true,scale=0.4]{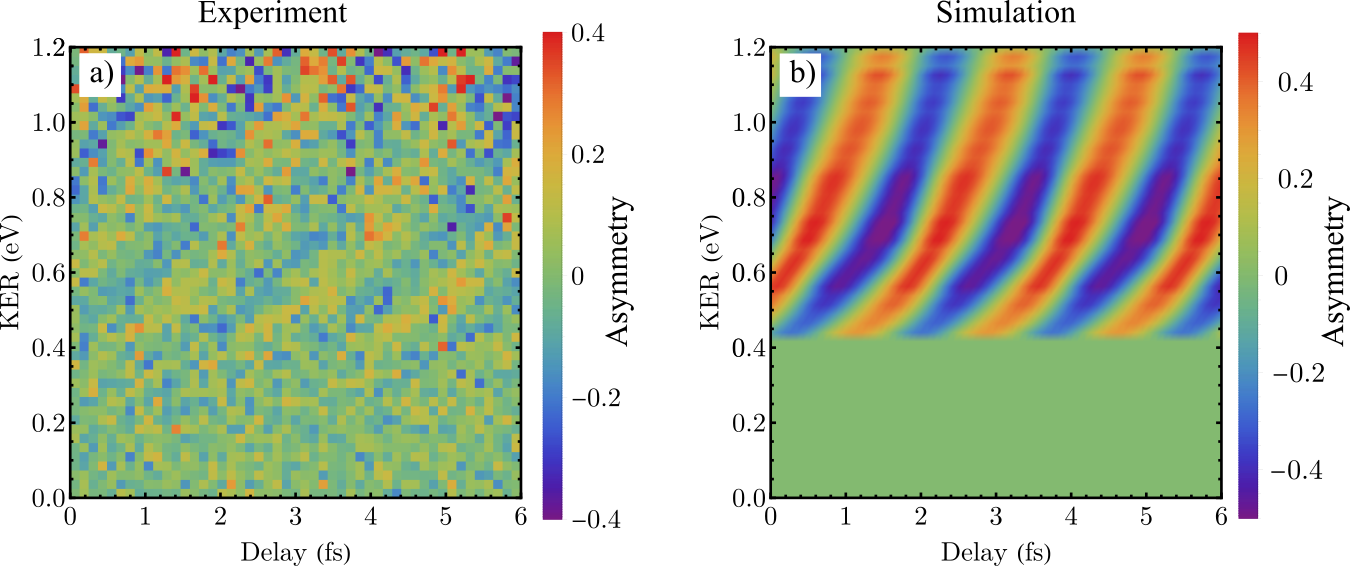}
\caption{\textbf{Time-dependent asymmetry parameter for OB: experiment vs. simulation.} The asymmetry parameter $A$ as a function of the time-delay between XUV and IR pulses. Comparison between the experiment a) \textcolor{red}{and simulation based on the WKB approximation b)}. {For the simulation, we plot Eq. 6 with retrieved $\alpha_o$ and $\beta_o$}.}
\label{fig:asymtimedep}
\end{figure}
\begin{figure}
\centering 
\includegraphics[keepaspectratio=true,scale=1.6]{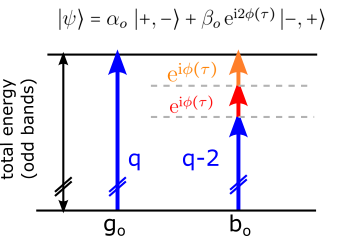}
\caption{Schematic illustration of the generation of Bell-like states using a combination of high-(blue arrows) and low-frequency photons (red and orange arrows) from the ground state of molecular hydrogen. The distribution of photons among the different excitation pathways determines the generation of a specific Bell-like state.}
\label{fig:PathsSimpl}
\end{figure}

{Identifying the main contributing pathways allows us to model and simulate the time- and KER-dependent asymmetry.
Two of three dominant pathways leading to OBs are illustrated in Fig.\ref{fig:PathsSimpl}. 
We discuss the meaning of the third path in the supplementary and safely ignore this path here for simplicity.}
Fig. \ref{fig:PathsSimpl} shows two energy states, which represents the initial state, the H$_2$ molecular ground state, and the final state, which includes the energy of the KER and the kinetic energy of the photoelectron.
Both contributions are present: pathway 1 (or $g_{o}$) belongs to the ground state dissociation and pathway 2 (or $b_{o}$) to bond softening dissociation.
{The color code in this Fig. \ref{fig:PathsSimpl} indicates the different function: the blue arrows illustrate two neighbouring XUV photons "$q$" and "$q-2$" that lead to photoionisation as explained in Fig. \ref{fig:h2paths}, the red arrow indicates that one IR photon is absorbed by the freed photoelectron after photoionisaion and the orange arrow indicates the IR photon absorbed by the molecular ion, see Fig. \ref{fig:h2paths}.}
Identifying these paths, allows to express the complex amplitudes in Eq. \ref{eq:bellv2} as $\alpha_o=g_{o}$ and $\beta_o=b_{o}\rm{e^{-\rm{i}2\phi}}$ (see supplement) and Eq. 3 can be rewritten in terms of the new coefficients by using an appropriate base according to \cite{Fischer2013}:
\begin{equation}
A=\frac{-2|g_{o}||b_{o}| \text{cos}\big(\Delta \phi_{go,bo}-2\phi(\tau)\big)}{|g_{o}|^2+|b_{o}|^2} ,  
\end{equation}
%\begin{equation}
%A=A(\alpha,\beta,\phi(\tau)) ,  
%\end{equation}  
with $\Delta \phi_{go,bo}=\text{arg}[g_o]-\text{arg}[b_o]$.
This shows that the asymmetry parameter not only depends on amplitudes and the relative phase shifts between the different paths, but also on the time-delay which can be controlled in our experiment.  
The phase terms can be obtained by perturbation theory and the Wentzel–Kramers–Brillouin (WKB) approximation.
The amplitudes are retrieved directly from the experimental data (see supplementary material). 
{The intensity of the NIR field controls the ratio of the relative amplitudes.  
At higher NIR field strengths, additional (multi-photon) paths could contribute to the signals, opening additional routes for control.}
The experimental results are very well reproduced (compare Fig. \ref{fig:asymtimedep}a) and b)) with this model and the numerical simulation.
{This confirms that the presented time-controlled few-photon ionisation and dissociation process preserves the natural, ubiquitous entanglement of the two-electron system in the ground state of the $H_2$ molecule.}

\vspace{10pt}
%\textcolor{gray}{Conclusion -- gray text will be removed later}\\
In conclusion, we have presented a proof-of-principle {demonstration of a sub-femtosecond control of the phase between entangled states} in molecular hydrogen by few-photon dissociation and coincident detection of all participating particles.
It will be interesting to apply this scheme to larger molecules or even solids, and thereby test quantum-dynamical theories describing effective decoherence effects arising e.g. due to coupling to complex electronic or internuclar/phononic degrees of freedom. 
It should be noted that for suitably chosen quantum systems, even visible frequencies are sufficient to implement the same ultrafast control scheme.

\newpage

\noindent{\Large \textbf{Methods}}

\noindent{\textbf{Experimental methods }}
  
The output of a linearly-polarized fiber laser  with a central wavelength of \unit[1030]{nm}, a pulse energy of \unit[1]{mJ}, and a pulse duration of \unit[50]{fs} at a repetition rate of \unit[50]{kHz} is divided into two beams with a 85/15 beam splitter and fed into an interferometer with a Mach-Zehnder configuration.

The main 85 percent (pump arm) is focused into an argon gas jet for high-harmonic generation (HHG) with a plano-convex lens with a focal length of \unit[500]{mm}. 
A \unit[200]{$\mu$m} aluminum foil is used to filter the incoming infrared beam as well as the lower harmonics   which results in the  XUV spectrum shown schematically in Fig. S2 \textbf{c} in the supplementary material. In the main text only a small region of the spectrum is shown for illustration purposes. 

The remaining 15 percent (probe arm) is delayed in time by changing the length of the probe arm of the interferometer using a controllable piezoelectric linear translation stage. An iris is  used to reduced the IR probe intensity to roughly $2\times 10^{11}$ $\mbox{W/cm}^2$ in order to avoid higher order photon-induced transitions during the experiment. 

Both arms are recombined and  focused using a grazing-incidence 2f-2f toroidal mirror into a supersonic gas jet of randomly-oriented cooled $H_2$ molecules inside a reaction microscope (cold target recoil ion momentum spectroscopy (COLTRIMS))\cite{DORNER200095} where the background pressure is kept below $2\times 10^{-10}$ mbar. Both ions and electrons are guided towards position-sensitive detectors with the help of an electric field. A homogeneous magnetic field is also used to reach a $4\pi$-electron-detection efficiency.

Momentum distributions of ions and electrons are obtained using the time of flight and the hit position on the detectors resulting in 3D momentum distributions. 
Three-particle momentum conservation is used to calculate the momentum vector of the missing H-atom. This way, the full kinematic information about the dissociation channels is obtained for each event.
During the measurements, we simultaneously monitor the two-particle break-up of $H_2$ into a molecular ion $H_2^+$ and a photoelectron. 
From this, we obtain in-situ information about the combined resolution of ions and electrons, which is in the order of 0.2 a.u. (atomic units) for ions and about 0.1 a.u. or better for electrons. 
These values have to be compared with typical total momenta of more than $p = 5\,$a.u. for the emitted H-atom (or proton) in case of dissociation. 
From these numbers, we conclude that the momentum of the undetected H-atom is reconstructed with an uncertainty of $\Delta p < 0.5$ a.u. or, in other words, with a relative accuracy of better than 10\% for $\Delta p/p$. 
For an overview of the setup, see Fig. S1 in the supplementary material.

 \newpage

 \noindent \textbf{Acknowledgments}
  The authors acknowledge and thank A. Buchleitner and E. Brunner for useful discussions. We further thank D. Bakucz Canário for helpful discussions and comments on the manuscript.
 
 \noindent \textbf{Funding} Max-Planck Society.

 \noindent \textbf{Author contributions} F.S. designed and constructed the pump-probe XUV-IR beamline and performed measurements and data analysis. P.F. and A.H. developed the theoretical description and numerical calculations. H.S. and A.H. contributed to the design and setting up of the beamline. R.M. and T.P. conceived the idea and conceptual interpretation. All authors discussed the results and contributed to the preparation of the manuscript.
 
\noindent \textbf{Competing interests} The authors declare no competing interests. 
\noindent \textbf{Data availability} Source data are provided in this paper. 
All other data that support the findings of this paper are available from the corresponding authors upon request.

\clearpage

% Include your paper's title here
\title{\smaller{}supplementary material for\\
\textbf{
{Sub-femtosecond optical control of entangled states}\\ 
%- \\ \textcolor{red}{A consequence of molecular entanglement}
}}

\author
{Farshad Shobeiry$^{ \ast,1}$, 
Patrick Fross$^{1}$, 
Hemkumar Srinivas$^{1}$,\\  
Thomas Pfeifer$^{ \ast,1}$, 
Robert Moshammer$^{ \ast,1}$,   
Anne Harth$^{ \ast,1,2}$  \\
\\
\normalsize{1 Max Planck Institute for Nuclear Physics, Saupfercheckweg 1, 69117 Heidelberg}\\
\normalsize{2 Center for Optical Technologies, Aalen University, Anton Huber Straße 1, 73430 Aalen}
\\
\\
\\ \normalsize{$^\ast$Corresponding authors. Email: shobeiry@mpi-hd.mpg.de, anne.harth@hs-aalen.de,}
\\ \normalsize{ thomas.pfeifer@mpi-hd.mpg.de, r.moshammer@mpi-hd.mpg.de. }
}
\setlength {\marginparwidth }{2cm} 

\date{}

\pagenumbering{roman}
\cleardoublepage
\pagenumbering{arabic}
\setlength{\parskip}{1em}
\baselineskip24pt

\maketitle

\section{Experimental setup}
\begin{figure}[H]
\centering 
\includegraphics[keepaspectratio=true,scale=0.4]{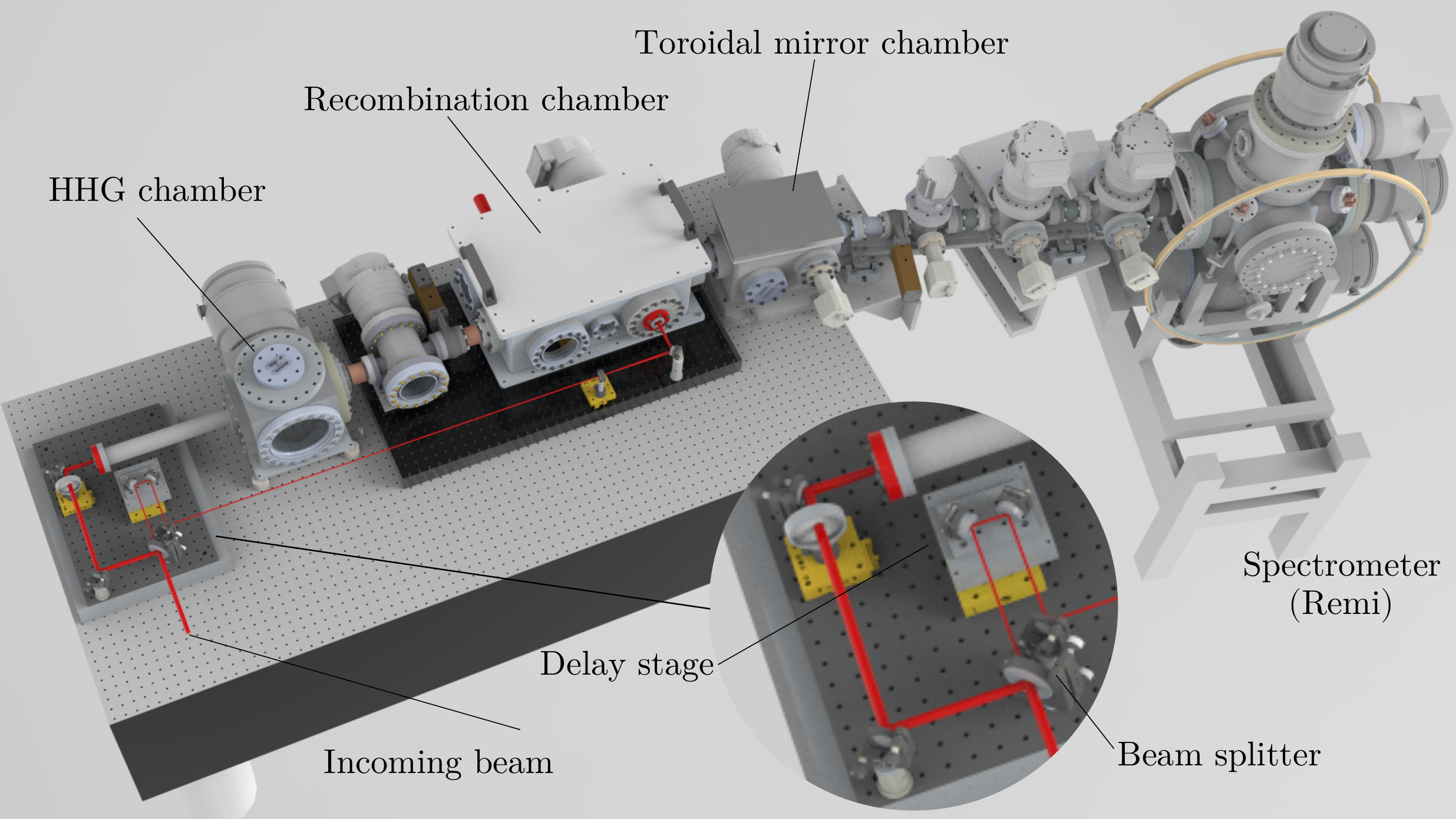}
\caption{{Overview of the attosecond beamline in Heidelberg.}}\label{setup}
\end{figure}
The measurement was performed with the attosecond beamline (Fig. S\ref{setup}) at the Max Planck institute for nuclear physics in Heidelberg. 
The incoming laser beam is divided into two beams using an 85/15 beam splitter. 
The main arm (85 percent of the beam) is focused into a gas jet for high-harmonic generation in the HHG chamber. The remaining 15 percent (probe arm) is delayed in time by changing the optical path length of the probe arm using a controllable linear translation state. The two arms are recombined in a recombination chamber and focused with a toroidal mirror into the reaction microscope (REMI) \cite{0034-4885-66-9-203}.
The interferometer is described in the Methods section of the main text. 

\subsection{Momentum resolution of the REMI}  
\begin{figure}
\centering 
\includegraphics[keepaspectratio=true,scale=1.2]{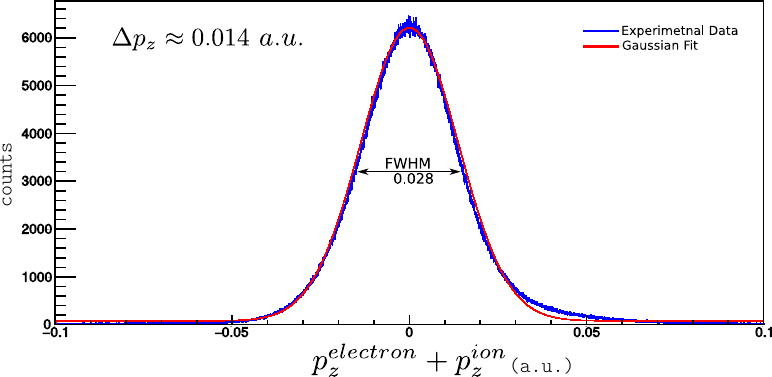}
\caption{Sum of the longitudinal momentum components of electrons (e$^-$) and ions (H$_2^+$) in the single ionization of H$_2$. Figure taken from \cite{shobeiry_2021}. }\label{momsumz}
\end{figure}

\begin{figure}
\centering 
\includegraphics[keepaspectratio=true,scale=1.15]{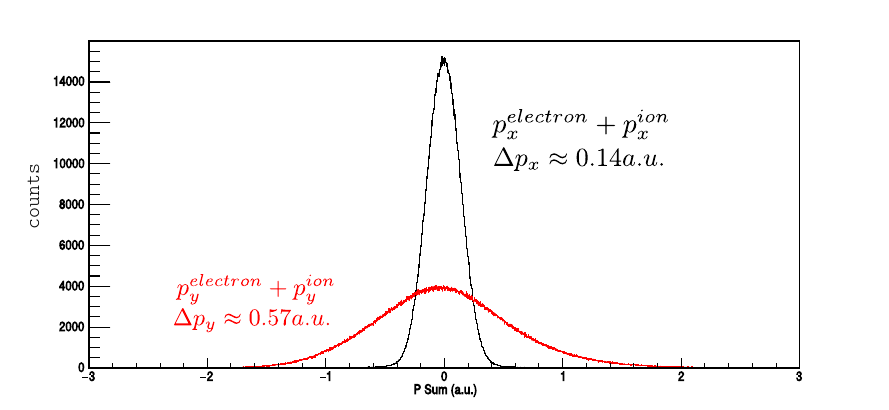}
\caption{Momentum sum for the two transversal components of electrons (e$^-$) and ions (H$_2^+$) in the single ionization of H$_2$. Figure taken from \cite{shobeiry_2021}.}  \label{momsumxy}
\end{figure}

Due to its geometry, the reaction microscope has a slightly different momentum resolution for different momentum components. 
The maximum resolution is along the spectrometer axis (longitudinal axis, z component). 
Fig. S\ref{momsumz} and Fig. S\ref{momsumxy} show the sum of ion-electron (H$_2^+$-electron coincidence) momenta for the single-ionization channel in H$_2$. 
The width of the distributions is the measure of the momentum resolution.
A dissociation event is determined by detecting an electron (e$^-$) and a proton (H$^+$) in coincidence. 
We recorded the data with a rate of 15 protons per second and detected in total over 2.1 millions $H^+$ events in coincidence with electrons.  

\subsection{Reconstruction of the molecular frame}
\begin{figure}
\centering 
\includegraphics[keepaspectratio=true,scale=0.5]{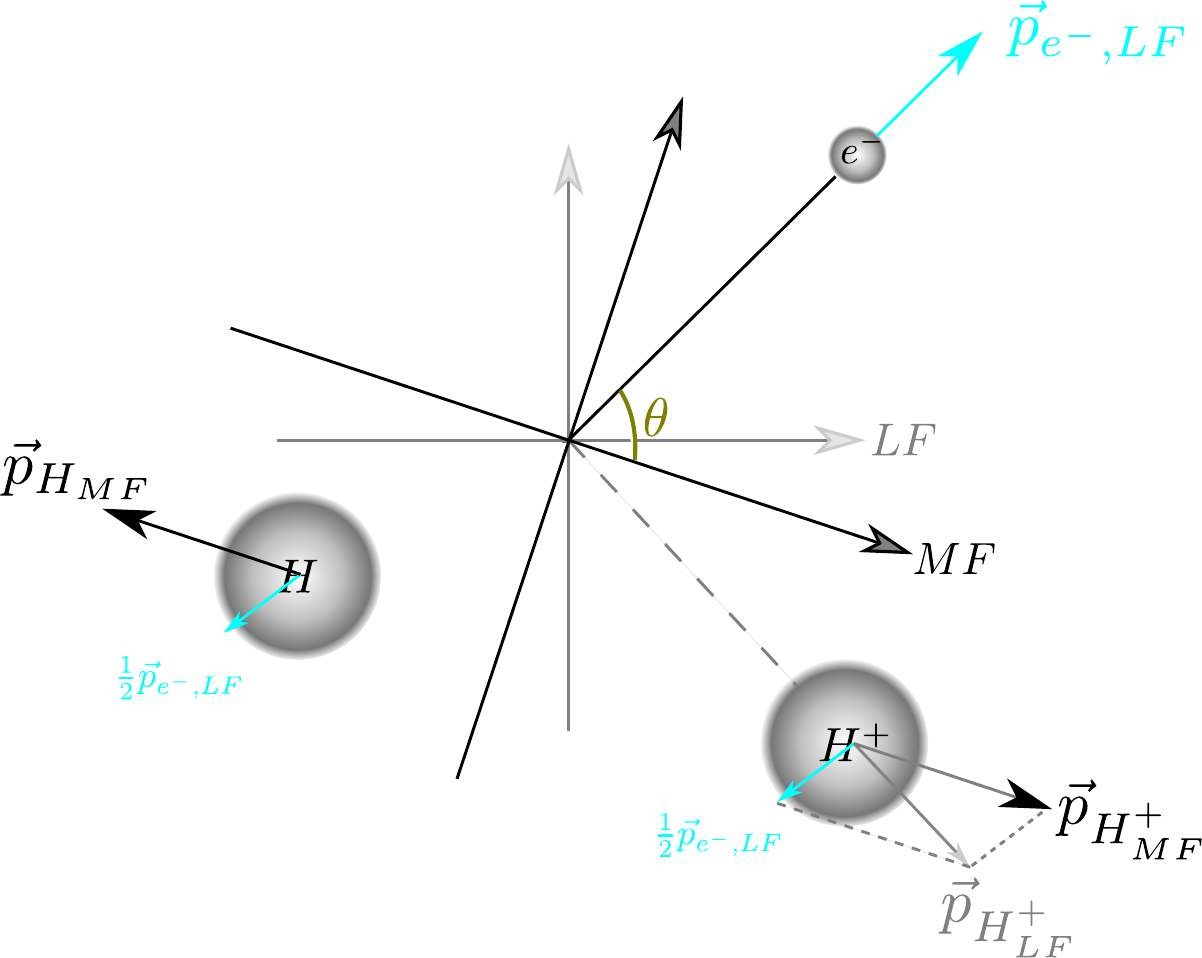}
\caption{Schematic representation of the reconstruction of the molecular frame. Figure taken from \cite{shobeiry_2021}.}  \label{mf}
\end{figure}

The detected recoil direction is the same as the dissociation direction at the moment of the interaction between the photon with the molecule based on the axial recoil approximation \cite{zare}. 
Initially, all detected fragments are in the laboratory frame (LF) of reference.  For photodissociation of H$_2$, the molecular frame (MF) momentum of the proton (see also Fig. S\ref{mf}) is given by:
\begin{equation}
\hat{\textbf{p}}^{H^+}_{MF}=\hat{\textbf{p}}^{H^+}_{LF}+  \frac{1}{2 }  \hat{\textbf{p}}^{e^-}_{LF},
\end{equation}
where $\hat{\textbf{p}}^{H^+}_{LF}$ and $\hat{\textbf{p}}^{e^-}_{LF}$ are the retrieved momentum vectors in the LF for the proton and electron, respectively.  
The neutral fragment of the dissociation (H) is not detected during our measurement. 
However, using momentum conservation, we can reconstruct the momentum vectors of H in post-analysis. 
The momentum of the incoming photon is neglected, as it is much smaller than the momenta of the dissociation fragments and the spectrometer momentum resolution. 
For instance, the momentum of a photon with the energy of \unit[30]{eV} is only \unit[0.008]{a.u.} ( \unit[1]{a.u.}of momentum is equivalent to \unit[1.995$\times 10^{-24}$]{kg.m.s$^{-1}$}) which is smaller than the momentum resolution of our spectrometer.

\section{Dissociation of H\texorpdfstring{$_2$}{Lg}}

\begin{figure}
\centering 
\includegraphics[keepaspectratio=true,scale=0.35]{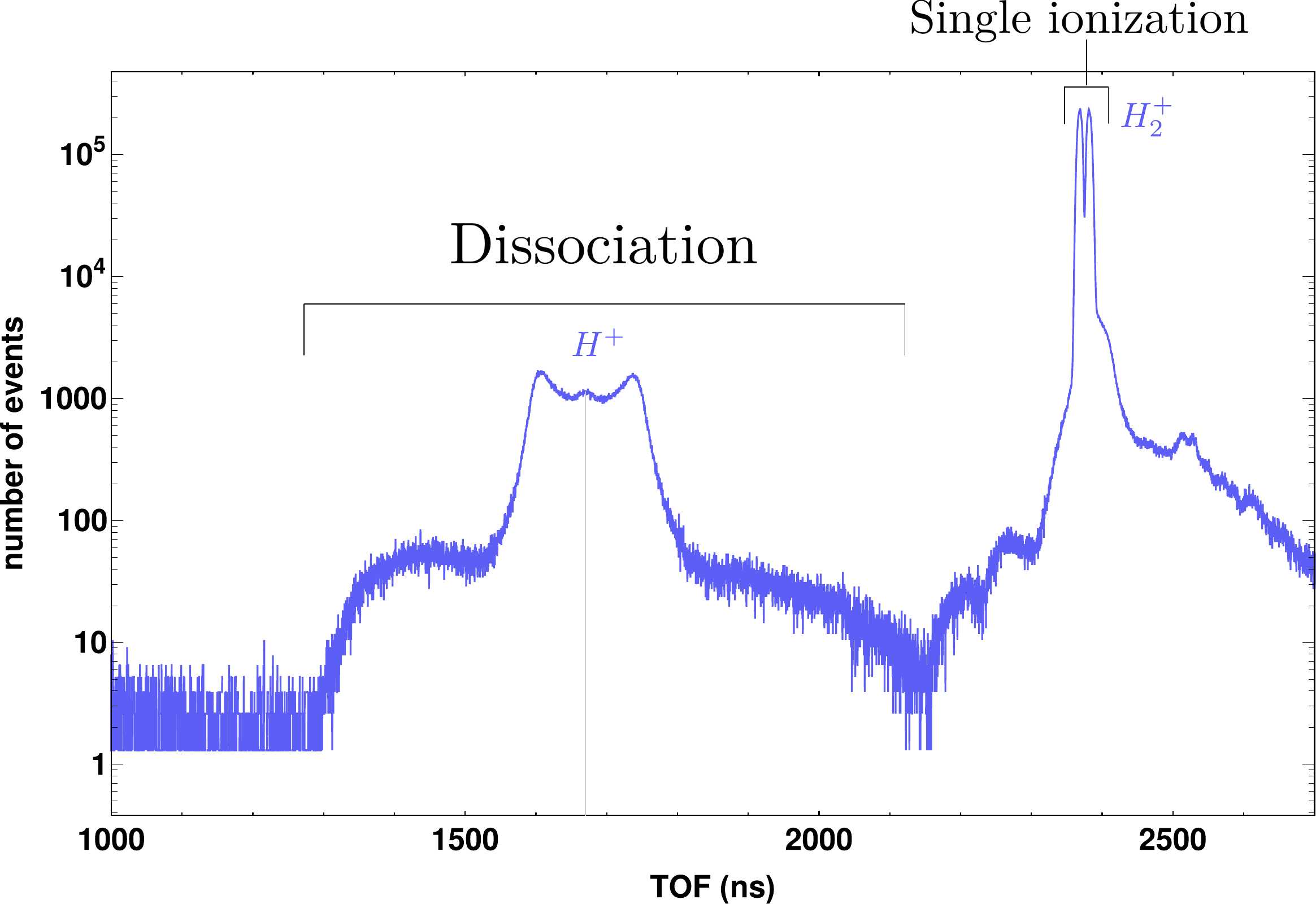}
\caption{Time of flight spectrum of detected ions as a result of the interaction between a combination of XUV and IR pulses with H$_2$ molecules.}\label{tofxuvir}
\end{figure}
Fig. S\ref{tofxuvir} shows a typical time-of-flight (TOF) spectrum of the detected ions as a result of ionization of H$_2$ with XUV and IR photons. 
These data show that the dominant interaction product is the single-ionization of the molecule.

\subsection{Interaction: XUV only }
Upon the absorption of XUV photons with energies higher than the dissociation limit of H$_2$ (I$_d$=18.1 eV) the molecule can be ionized leading consecutively to the dissociation of the molecule.
 \begin{figure*}
\centering 
\includegraphics[keepaspectratio=true,scale=1.0 ]{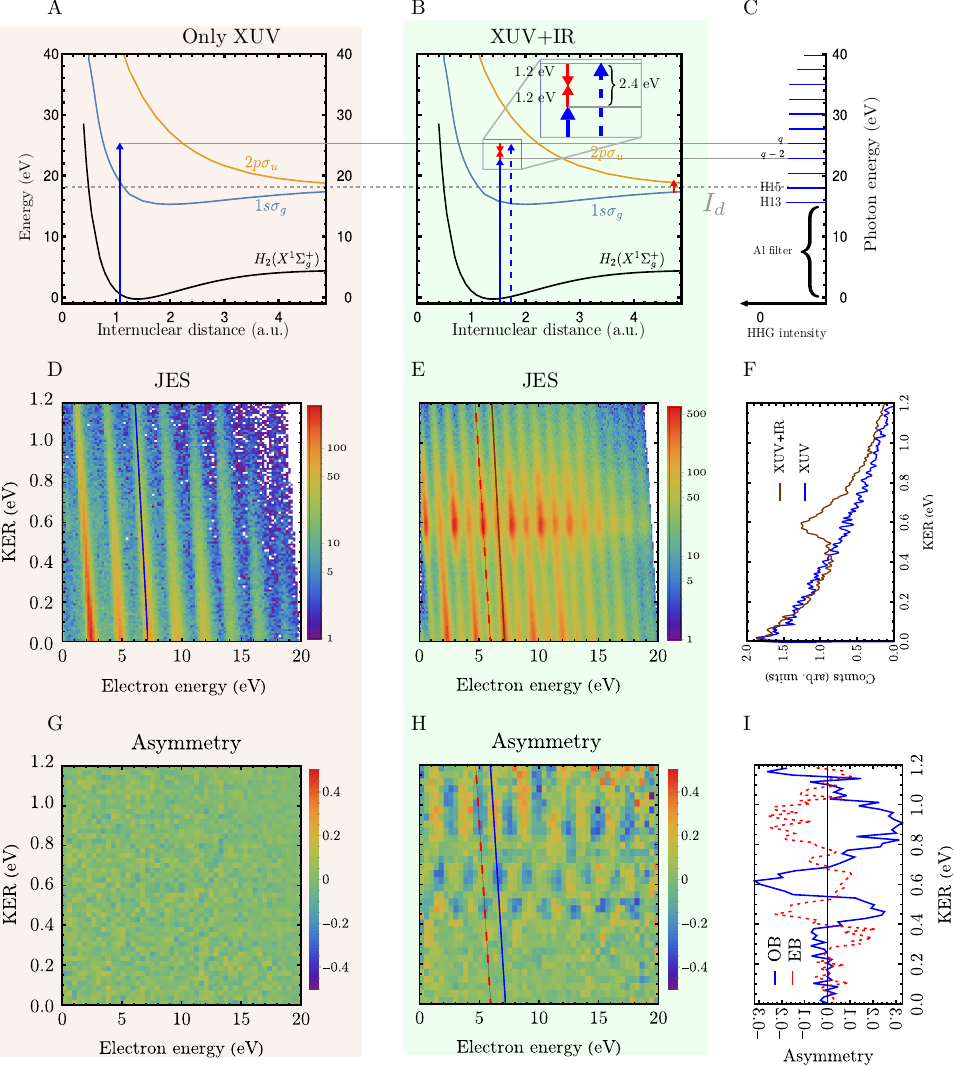}
\caption{Dissociation of H\texorpdfstring{$_2$}{Lg}: The relevant potential energy curves of H$_2$ are given in A and B. A: schematic interaction of the 21st harmonic only. B: schematic interaction of the XUV and IR photons. C: schematic spectrum of the XUV radiation resulting from high order harmonic generation. Lower part of the spectrum is filtered out by an aluminum filter. D: JES for the dissociation of H$_2$ with XUV light only. E:  JES for the dissociation of H$_2$ with XUV+IR pulses. F: projection of the OB-21 created from XUV only (blue curve) and XUV+IR (brown curve) interaction onto the KER axis. G: asymmetry parameter $A$ for XUV only. H: $A$ for XUV+IR Pulses. I: $A$ for EB 20 and OB 21.}
\label{viblevel}
\end{figure*}
Fig. S\ref{viblevel}-A shows the schematic interaction of an XUV photon with H$_2$. 
The dissociation limit is marked with a horizontal dashed line at 18.1 eV.
This photon belongs to the 21th harmonic of the XUV spectrum with an energy of 25.2 eV. 
A schematic representation of the high order harmonic spectrum is shown in Fig. S\ref{viblevel}-C.
As a result, the molecule dissociates along the ionic state 1s$\sigma_g$ (ground-state dissociation (GS)). 
GS is the dominant dissociative channel when applying XUV photon energies up to \unit[27]{eV}\cite{doi:10.1063/1.465352}. 
GS leaves the ionic fragment with an energy of less than \unit[1]{eV}. 
The corresponding events from dissociation with harmonic 21st appear on a diagonal line marked with a solid blue line in the joint-energy spectrum (JES) histogram in Fig. S\ref{viblevel}-D.  
The other lines in Fig. S\ref{viblevel}-D correspond to other harmonics of the XUV spectrum. 
We call these lines "odd bands" (OBs) since an odd number of photons (in this case only one XUV photon) is absorbed. 
The projection of the marked band onto the KER axis is shown by the blue curve in Fig. S\ref{viblevel}-F.

%The ratio $\frac{H^+}{H_2^+}$ is almost 2 $\%$ which is in a good agreement with cyclotron experiments \cite{Browning_1973}.
\subsection{Interaction: XUV + IR }
The schematic interaction of a combination of  XUV and weak IR photons with H$_2$ is shown in Fig. S\ref{viblevel}-B. 

In this case, one IR photon (red arrows) is absorbed or emitted (shown in the inset of panel B) by the photoelectron which results in events on diagonal lines (bands) between the OBs. 
One example is marked with a dashed red line in Fig. S\ref{viblevel}-E which we called an "even band" (EB) as a result of the absorption of e.g. an XUV photon from harmonic 19  plus an IR photon and/or the absorption of an XUV photon from harmonic 21  minus an IR photon.
 
In general, an additional IR photon can also be absorbed by the molecular ion due to bond softening at an  internuclear distance (R) of around 5 a.u., where the 1s$\sigma_g$ and 2p$\sigma_u$ curves are energetically one IR photon apart (this process is indicated by the red arrow in Fig. S\ref{viblevel}-B at R $\approx$ 5 a.u.). This process results in an enhancement in the dissociation rate and is called bond softening (BS) dissociation.

The projection of the previously selected OB (this time marked with the brown solid line) in the case of the interaction with an XUV and an IR photon is shown by the brown curve in panel Fig. S\ref{viblevel}-F.
We normalized the two curves at the KER around zero to emphasize the difference at a KER of around 0.6 eV. 
Note that, in this case, the region at  KER $\approx$ 0.6 eV can be reached by the interaction with one XUV photon (from harmonic 21) only (GS) as well as by the interaction with one XUV photon from the next lower harmonic (harmonic 19) plus two IR photons (one absorbed by the photoelectron and the other by the molecular ion (BS)) or by one XUV photon from the same 21st harmonic followed by the emission of an IR photon by the photoelectron and absorption of another IR photon by the molecular ion (BS).
(Note: The latter is a "third path" (see also Fig. S\ref{SF:allQP}), omitted in the main text for simplicity.)  

\subsection{Asymmetry parameter }  
The experimental results for the asymmetry parameter $A$ are shown in Fig. S\ref{viblevel}-G and -H. 
Panel G shows that $A$ is zero in the case of dissociation triggered only with XUV photons. 
This is a clear indication that the asymmetric photoelectron emission only takes place when different dissociation pathways (leading to the same final state) interfere.
However, in Fig. S\ref{viblevel}-H we observe asymmetric photoelectron emission in the KER region where pathways contributing to BS and GS dissociation overlap (KER $\approx$ 0.6eV, see Fig. S\ref{viblevel}-F). 
An important feature in Fig. S\ref{viblevel}-H is that all OBs show the same trend. 
The same is true for EBs, all EBs show the same trend. Another feature is that the OBs and EBs are $\pi$ out of phase due to the extra absorbed photon in the EBs.
This feature is underlined in the projection of an odd and even band onto the KER axis in Fig. S\ref{viblevel}-I.

\subsection{Pathways in even bands}
In the discussion above, possible dissociation pathways according to the lowest order perturbation are described for the OBs. 
In the case of EBs (see e.g. Fig. S\ref{SF:allQP}),  an XUV photon with an energy higher than I$_d$    leads to dissociation through 1s$\sigma_g$ with KERs<2 eV. 
The photoelectron emits an IR photon leaving  the electron energy   $E_e=\gamma_q-I_d-KER-\hbar\omega$ and the parity of the bound and continuum electron  becomes  gerade $\ket{+}$. For the pathway including bond softening, the molecule is ionized with the next lower harmonic into the ground state of the molecular ion 1s$\sigma_g$. 
The photoelectron energy is in this case $E_e=\gamma_{q-2}  -E_b$ with $E_b$ being the energy of the bound vibrational level in the ground state of the molecular ion. 
The parity of the photoelectron is ungerade $\ket{-}$. 
The molecular ion which contains the bound electron, absorbs   an IR photon which promotes the molecular ion to the repulsive 2p$\sigma_g$ curve. 
The molecule eventually dissociates, leaving the bound electron with ungerade parity  $\ket{-}$.

\subsection{Electron emission asymmetry}
In this section, we establish a connection between the asymmetry parameter $A$ (main text Eq. 3) and the coherent superposition of the dissociative pathways.

Since we have more than one pathway leading to the final state and these pathways are of two different natures - namely: ground-state (GS) dissociation and bond softening (BS) dissociation - we write the final state in a general manner in the form:  

\begin{equation}
\ket{\psi_{ob}  }=  \boldsymbol{c}_{gs} \ket{+,-}  +\boldsymbol{c}_{bs} \ket{-,+}   ,  \label{cohsup1}
\end{equation} 
for OBs and 

\begin{equation}
\ket{\psi_{eb}  }=  \boldsymbol{c}_{gs} \ket{+,+}  +\boldsymbol{c}_{bs} \ket{-,-}   ,  \label{cohsupeven}
\end{equation} 
for EBs, 
where $\boldsymbol{c}_{gs}$ and  $\boldsymbol{c}_{bs}$ are complex numbers. 
Both, the amplitudes and phases are relevant.

The superposition of molecular orbitals with different parities leads to the localization of a \textbf{bound electron} either on the left or right side of the nucleus. 
A similar spatial asymmetry occurs for a \textbf{photoelectron} placed in a superposition of spherical harmonics with either positive $\ket{+_e}$ or negative $\ket{-_e}$ parities (see \cite{Fischer_2013}).
 
\begin{equation*}
\includegraphics[keepaspectratio=true,scale=0.35]{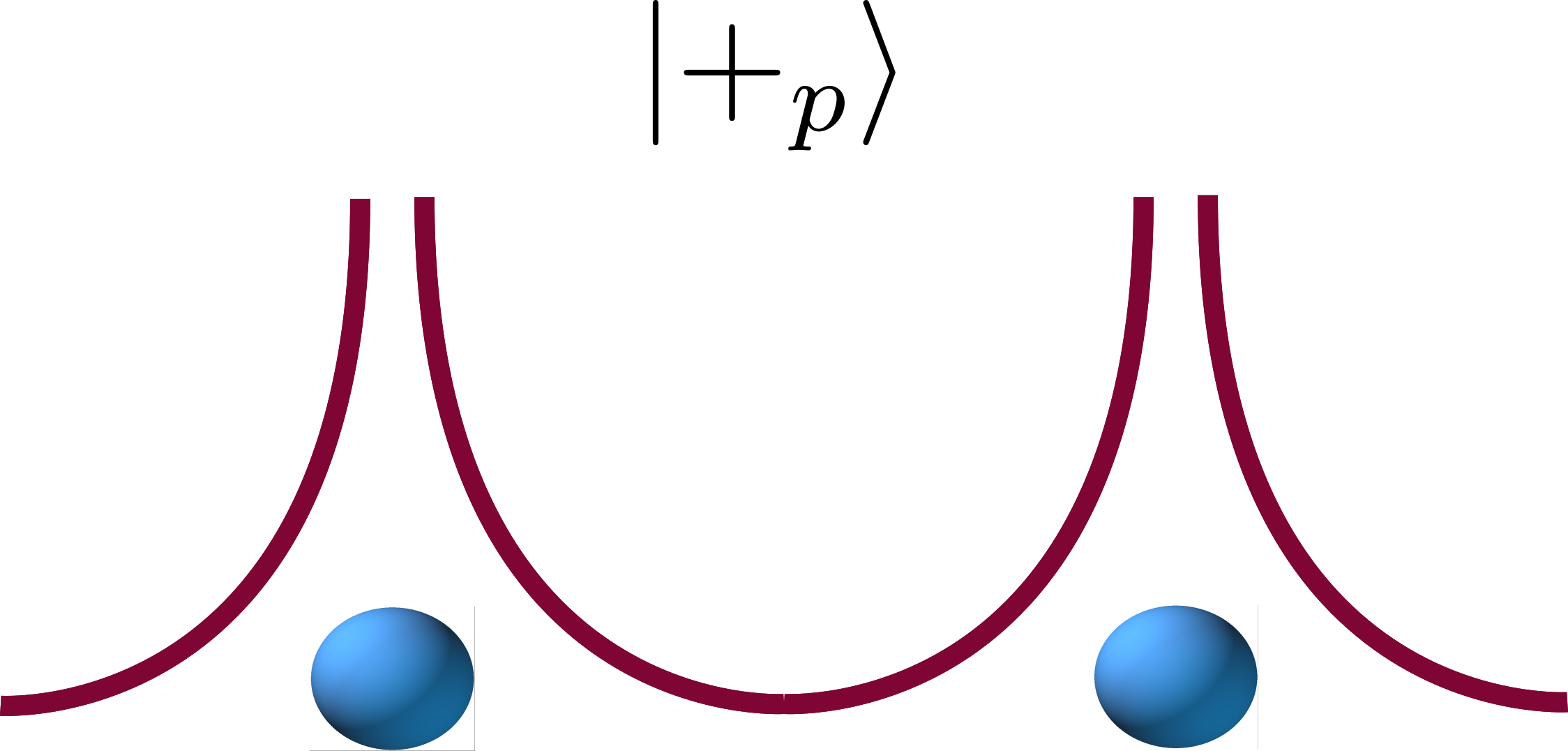}\; +\; \includegraphics[keepaspectratio=true,scale=0.35,trim=0 +3.2cm 0 0]{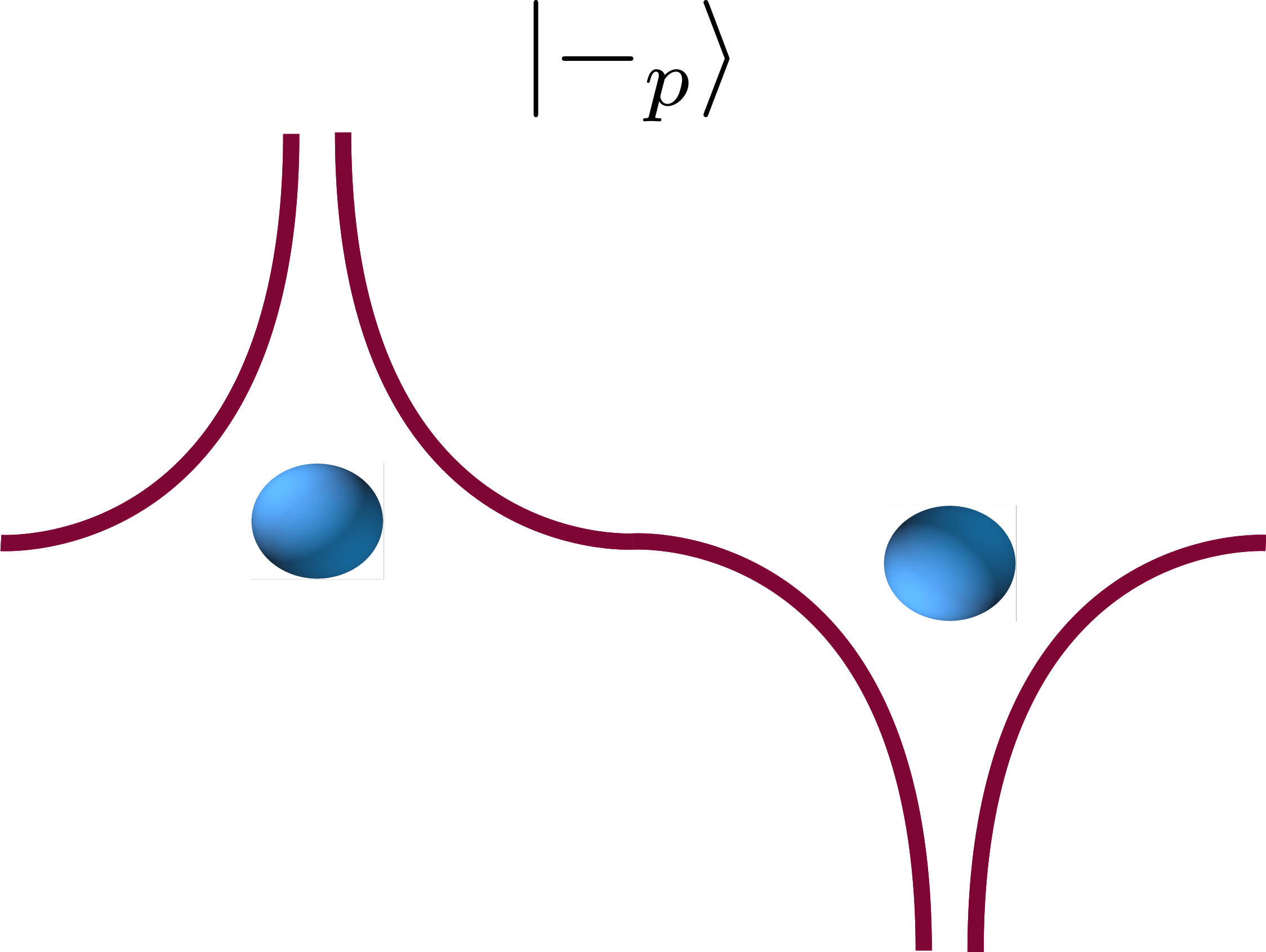}\; =\; \includegraphics[keepaspectratio=true,scale=0.35,trim=0 +1.5cm 0 0]{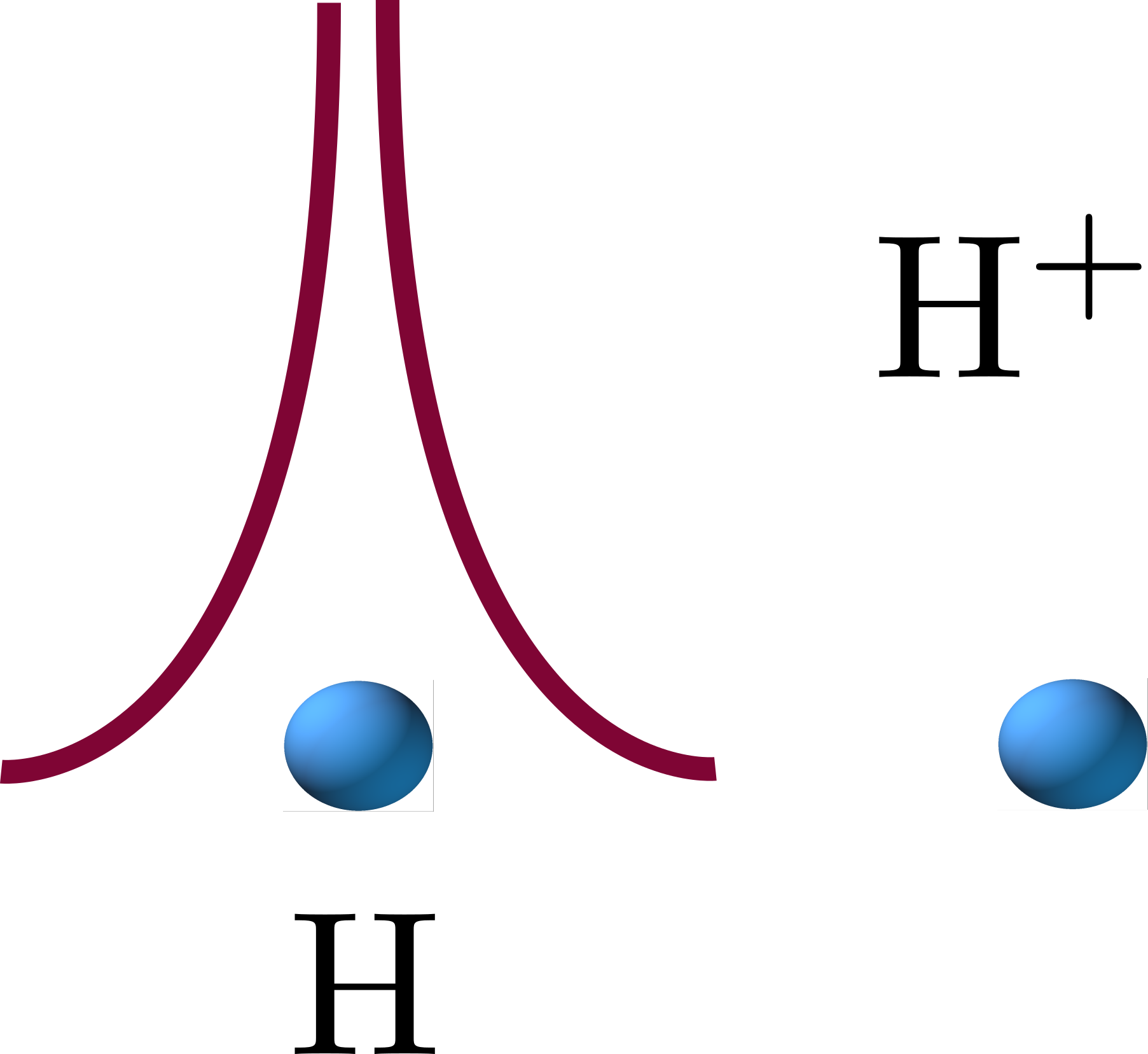}\label{wavefunctionaddition1}
\end{equation*}

\begin{equation*}
\includegraphics[keepaspectratio=true,scale=0.35]{gradestate.png} \;-\; \includegraphics[keepaspectratio=true,scale=0.35,trim=0 +3.2cm 0 0]{ungradestate.png}\; =\; \includegraphics[keepaspectratio=true,scale=0.35,trim=0 +1.5cm 0 0]{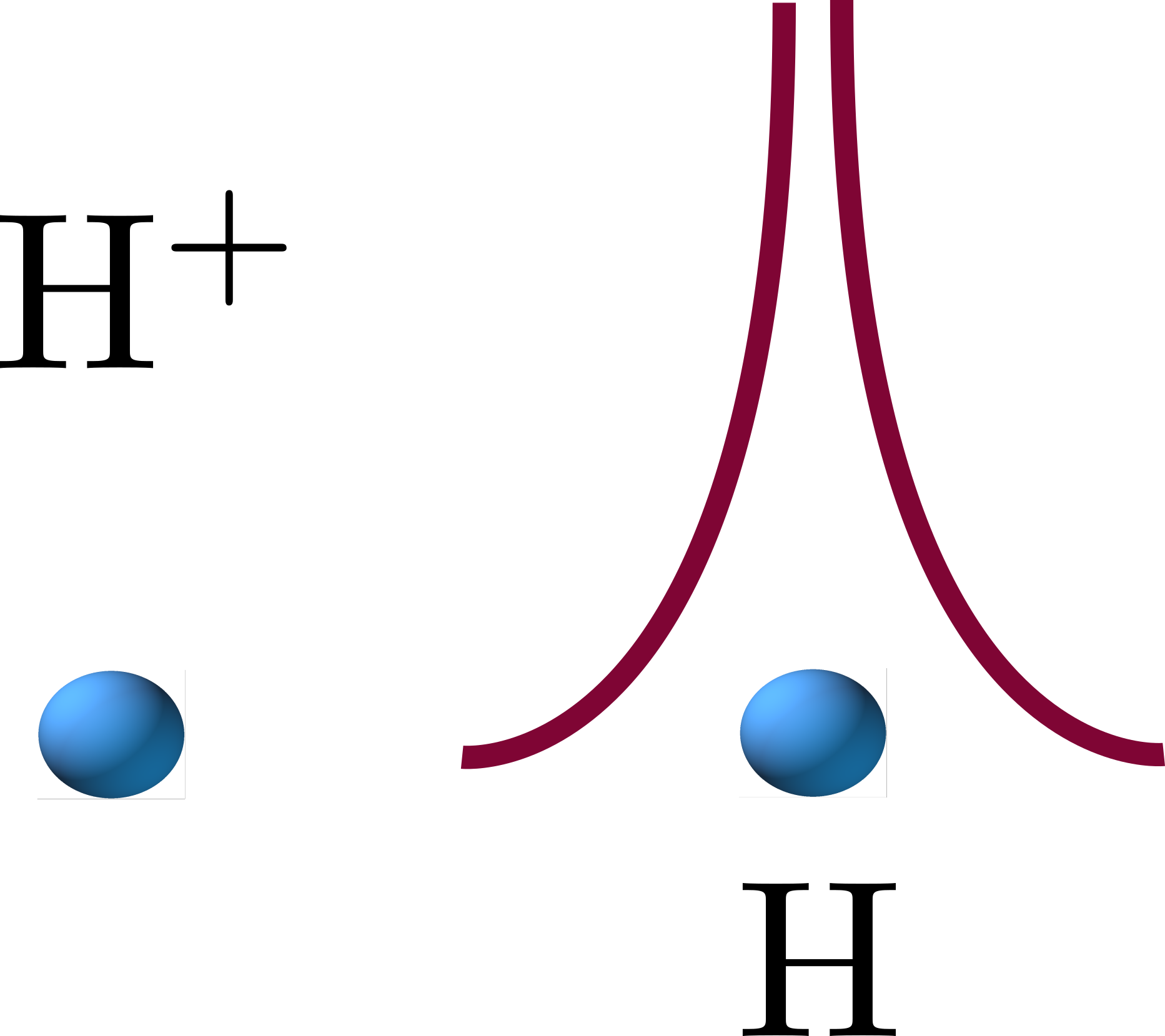}\label{wavefunctionaddition2}
\end{equation*}
\vspace{4mm}

As a result, we can now define a left-right basis as follows:  
\begin{equation*}
\begin{split}
\ket{H_{right}^+}&=\ket{+_p}+\ket{-_p},\\ \ket{H_{left}^+}&=\ket{+_p}-\ket{-_p},\\
\ket{e_{right}^-}&=\ket{+_e}+\ket{-_e},\\
\ket{e_{left}^-}&=\ket{+_e}-\ket{-_e}.
\end{split}
\end{equation*}
This leads to the following four electron and proton combinations. 
Here exists two cases for both - the electron and the proton - being emitted in the same direction (and vice versa).
\begin{equation*}
\begin{split}
&\ket{\psi_{\theta<90}^1}=\ket{H_{right}^+}\otimes\ket{e_{right}^-},\qquad\qquad  \includegraphics[keepaspectratio=true,scale=0.1,trim=0 +6.5cm 0 0]{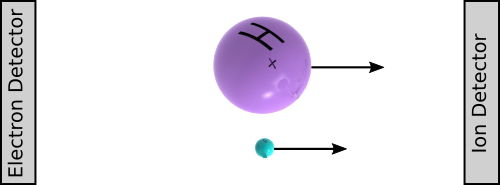}\\
\\
&\ket{\psi_{\theta<90}^2}=\ket{H_{left}^+}\otimes\ket{e_{left}^-},\;\;\;\qquad\qquad   \includegraphics[keepaspectratio=true,scale=0.1,trim=0 +6.5cm 0 0]{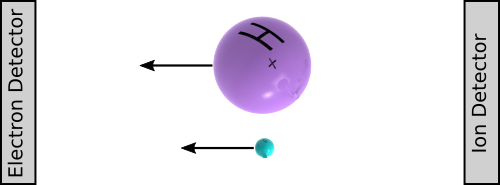}\\\\
&\ket{\psi_{\theta>90}^1}=\ket{H_{right}^+}\otimes\ket{e_{left}^-},\;  \qquad\qquad    \includegraphics[keepaspectratio=true,scale=0.1,trim=0 +6.5cm 0 0]{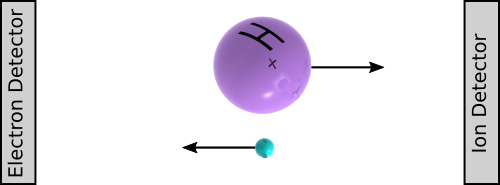}\\\\ 
&\ket{\psi_{\theta>90}^2}=\ket{H_{left}^+}\otimes \ket{e_{right}^-}, \;\qquad\qquad 	\includegraphics[keepaspectratio=true,scale=0.1,trim=0 +6.5cm 0 0]{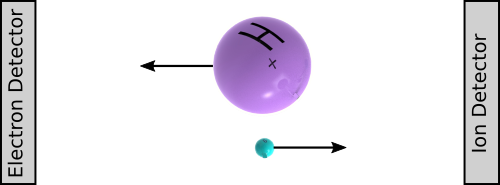} 
\end{split} 
\end{equation*}

\vspace*{25px}
\noindent $\theta$ is the angle between the photoelectron and the ejected proton (see Fig. S4 and Fig. 2 in the main text).
The transition coefficients $\boldsymbol{c}_{gs}$ and  $\boldsymbol{c}_{bs}$ can be calculated by projecting the final state (Eq.S.3, Eq.S3) onto different direction states using the above equations:
\begin{equation}
\begin{split}
\abs{\braket{\psi_{\theta<90}^1}{\psi }}^2 &=(c_{gs}+c_{bs})(c_{gs}^*+c_{bs}^*),\\
\abs{\braket{\psi_{\theta<90}^2}{\psi }}^2 &=(-c_{gs}-c_{bs})(-c_{gs}^*-c_{bs}^*),\\
\abs{\braket{\psi_{\theta>90}^1}{\psi }}^2 &=(-c_{gs}+c_{bs})(-c_{gs}^*+c_{bs}^*),\\
\abs{\braket{\psi_{\theta>90}^2}{\psi }}^2 &=(c_{gs}-c_{bs})(c_{gs}^*-c_{bs}^*).
\end{split}
\end{equation}
We define the number of dissociation events as:
\begin{equation}
\begin{split}
N_{\theta<90}&=\abs{\braket{\psi_{\theta<90}^1}{\psi_f}}^2+\abs{\braket{\psi_{\theta<90}^2}{\psi_f}}^2,\\
N_{\theta>90}&=\abs{\braket{\psi_{\theta>90}^1}{\psi_f}}^2+\abs{\braket{\psi_{\theta>90}^2}{\psi_f}}^2.
\end{split}
\end{equation}
Using now the definition of the asymmetry parameter $A$ (Eq.3 of the main text), we can rewrite $A$ as a function of $\boldsymbol{c}_{gs}$ and $\boldsymbol{c}_{bs}$:
\begin{equation}
\begin{split}\label{asymequ}
A&=-\frac{2\text{Re}[c_{gs}c_{bs}^*]}{\abs{c_{gs}}^2+\abs{c_{bs}}^2}\\
&=-\frac{2\abs{c_{gs}}\abs{c_{bs}}\text{cos}(\phi_{gs}-\phi_{bs} )}{\abs{c_{gs}}^2+\abs{c_{bs}}^2},
\end{split}
\end{equation}
where $\phi_{gs,bs}=\text{arg}[c_{gs},c_{bs}]$.

\section{Model based on perturbation theory and WKB approximation}
In this section we introduce a model that supports that the origin of the time-dependent asymmetry lies in the interference of photoelectrons coming from GS and BS dissociation quantum pathways where at least two neighboring harmonics in the XUV spectrum are involved.

\begin{figure}
\centering 
\includegraphics[keepaspectratio=true,scale=1.5]{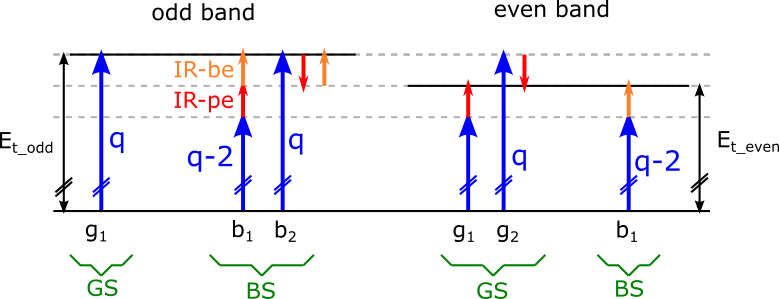}
\caption{\textbf{Quantum pathways for odd and even bands}. The blue arrows represent the XUV photon  with different discrete photon energies: e.g. $q-2$ is the next lower harmonic to $q$ with an energy difference of two IR photons. The red arrow represents an IR photon absorbed or emitted by the photoelectron (IR-pe), while the orange one represents an IR photon  absorbed by the bound electron (IR-be) leading to bond softening. } \label{SF:allQP}
\end{figure}

%\subsection{Model}
Many quantum pathways are involved in the experiment. However, for the GS and BS dissociation pathways, we consider only the lowest photon transitions due to vanishing intensities of both XUV and IR pulses. These lowest order pathways are shown in Fig. S\ref{SF:allQP}.
All these paths interfere and have a contribution to the final dissociation probability, and affect the asymmetry parameter $A$.
 
\subsection{Odd bands}
For now, we consider \textbf{odd bands} only. The following discussion is repeated for even bands in section 3.5. 
The complex coefficients are called $g_1$ for the GS pathway, $b_1$ and $b_2$ for the BS pathway according to Fig. S\ref{SF:allQP}.   
Eq.\ref{cohsup1} becomes:
\begin{equation}
\ket{\psi  }=  g_1 \ket{+,-}  +(b_1+b_2) \ket{-,+}. \label{cohsup2}
\end{equation} 
Note that  for pathway $g_1$ one photon and for pathways $ b_1$ and $b_2$ three photons are involved. 
With this, the dissociation probability reads:
\begin{equation}
P_{odd}= |g_{1}+ b_{1}+b_{2}|^2.
\end{equation}
%\begin{equation}
%P_{odd}=|g_1|^2+|b_{1}|^2+|b_{2}|^2+2(|g^{}_{1}||b^{}_{1}|\cos(\Delta\phi_{g_1,b_1})+|g^{}_{1}||b^{}_{2}|\cos(\Delta\phi_{g_1,b_2})+|b^{}_{1}||b^{}_{2}|\cos(\Delta\phi_{b_1,b_2})),\label{eq:oddProb3P}
%\end{equation}
The asymmetry parameter from Eq. \ref{asymequ} becomes: 
\begin{equation}
\begin{split}
    A_{odd}&=\frac{-(g_1(b_1+b_2)^{*}+(b_1+b_2)g_1^{*})}{|b_1+b_2|^2+|g_1|^2}\\
    &= \frac{-2(|g^{}_{1}||b^{}_{1}|\cos(\Delta\phi_{g_1,b_1})+|g^{}_{1}||b^{}_{2}|\cos(\Delta\phi_{g_1,b_2}))}{|g_1|^2+|b_{1}|^2+|b_{2}|^2 +2 |b^{}_{1}||b^{}_{2}|\cos(\Delta\phi_{b_1,b_2})}\label{eq:oddAsy3P},
\end{split}
\end{equation}
with $\Delta\phi_{bi,bj}= \arg[b_i]-\arg[b_j]$ and $\Delta\phi_{gi,bj}= \arg[g_i]-\arg[b_j]$. 

The complex coefficients correspond to the specific components of the dipole transition element from the vibronic ground state to the continua \cite{doi:10.1002/cphc.201200974}.
Within the Franck-Condon (FC) approximation, the electronic and the nuclear components of the total wave function (Eq.\ref{cohsup2}) can be separated $\psi(r,R)= \chi(R)\phi(r,R)$. 
We neglect the dependence of the electronic matrix elements on the nuclear position.
This leads to the complex expressions:
\begin{equation}
 g_1=\underbrace{FC_{gs}\;\mathrm{e}^{\mathrm{i}\Theta_{gs}}}_{\substack{nuclear\\contribution}}\times\underbrace{ M^{(1)}_{g1}}_{\substack{photo-e\\contribution}},
\end{equation}
\begin{equation} 
 b^{}_{1}=\underbrace{FC_{bs}\;\mathrm{e}^{\mathrm{i}\Theta_{bs}}}_{\substack{ nuclear\\contribution}}\underbrace{\textcolor{black}{\mathrm{e}^{\mathrm{i}\phi(\tau)}}}_{\substack{bound-e\\contribution}}\times\underbrace{ M^{(2)}_{b1}}_{\substack{photo-e\\contribution}},
\end{equation}
\begin{equation} 
  b^{}_2=\underbrace{FC_{bs}\;\mathrm{e}^{\mathrm{i}\Theta_{bs}}}_{\substack{ nuclear\\contribution}}\underbrace{\textcolor{black}{\mathrm{e}^{\mathrm{i}\phi(\tau)}}}_{\substack{bound-e\\contribution}}\times\underbrace{ M^{(2)}_{b2}}_{\substack{photo-e\\contribution}},
\end{equation}
where $FC$ stands for a contribution that represents the nuclear part. 
The nuclear part has a phase $\Theta_{gs/bs}$.
The contribution for the photoelectron is accounted for by the complex multi-photon matrix element $M^{(N)}$ where $N$ is the number of involved photon transitions.  
Bond softening is a photo-induced process. 
In order to account for the phase of this field, the phase of the IR field $e^{i\phi(\tau)}$ is multiplied to the bond softening matrix element in Eq.S11 and S12. 

\subsection{WKB-phase}
\begin{figure}
\centering 
\includegraphics[keepaspectratio=true,scale=0.9]{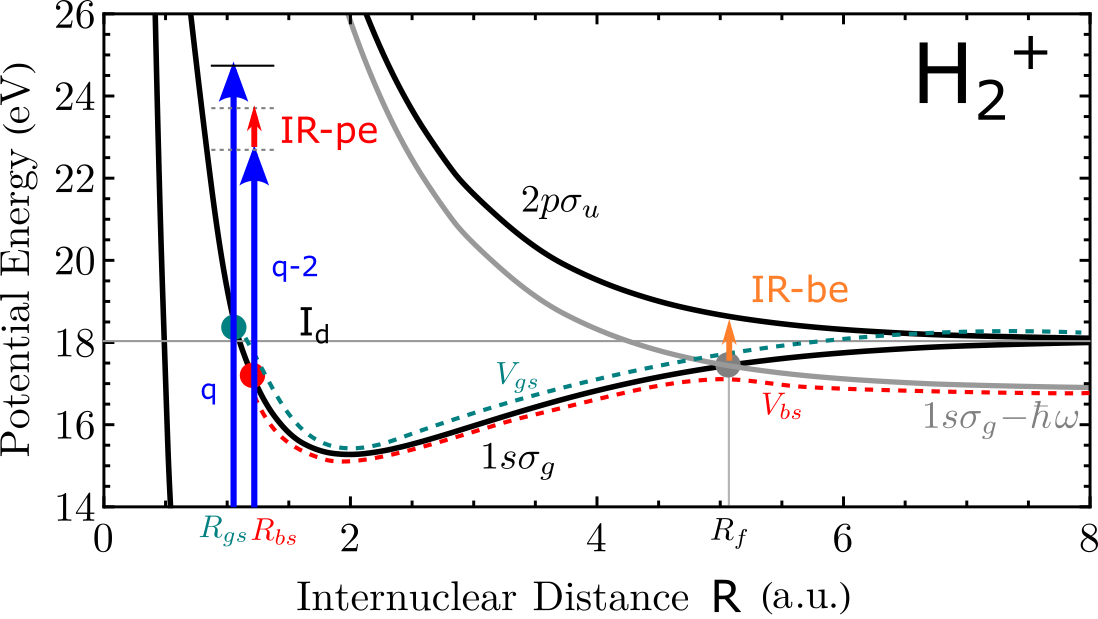}
\caption{Schematic representation of the quantum paths $g_1$ and $b_1$ in the $H_2$ molecular potential curves. These two quantum paths are the most significant ones  that lead to a time-dependent asymmetry in odd bands.}\label{SF:WKPpaths}
\end{figure}

The phases $\Theta_{gs/bs}$ account for the accumulated phases of the nuclei moving in the given potential energy curves $V_{gs}$ and $V_{bs}$, see Fig. S\ref{SF:WKPpaths}.
The phases from the moving nuclei are estimated with the WKB Wentzel\-–Kramers\-–Brillouin (WKB) 
%\cite{PhysRevLett.110.213002} 
approximation:
\begin{equation}
    \Theta_{gs}= \int_{R_{gs}}^{\infty} dR \sqrt{2\mu (KER+I_d-V_{gs}(R))},
\end{equation}
\begin{equation}
    \Theta_{bs}= \int_{R_{bs}}^{\infty} dR \sqrt{2\mu(KER+I_d-V_{bs}(R))},
\end{equation}
where the integrand is the nuclear momentum with $\mu$ being the reduced mass of H$_2^+$.

Please note the different limits of the integrals. 
For ground-state dissociation, the phase  $\Theta_{gs}$ is calculated by integrating from the internuclear distance  $R_{gs}$ to the limit of $R\rightarrow\infty$ where the molecule dissociates along the 1s$\sigma_g$ curve with a KER of 0.6 eV. 
The phase of the pathway ($\Theta_{bs}$) is obtained by the integral in equation S14 which  has a different lower limit since the process starts with the ionization of the molecule leaving the ion in the bound ground state. The integration pathway follows the laser-dressed potential curve (dashed red curve (V$_{bs}$)) at the limit of the adiabatic 1s$\sigma_g$-$\hbar\omega$ curve obtained by diagonalizing the diabatic potential matrix \cite{PhysRevA.48.2145}. The width of the avoided crossings at R$_f$ is obtained by taking into account the IR intensity during the measurement. 

\subsection{Phase of electric dipole matrix element}
For this model approach, the electric dipole matrix elements, $M_{}$, are approximated to be independent of $R$. 
According to the shown paths in Fig. S\ref{SF:allQP}, $M_{g1}^{(1)}$   describes a first-order perturbation process and $M_{b1}^{(2)},M_{b2}^{(2)}$  describe second-order perturbation   processes.
Although we deal with a real two-electron system, the single active electron approximation can be used \cite{PhysRevA.77.063403}, since the interaction with the light only changes the electronic state of the photo-electron; while the bound electron remains in the ground state of the molecular ion.

The complex electric dipole matrix elements (assuming only linearly polarized light) and their asymptotic phases can be written as \cite{PhysRevA.103.022834}:
%\begin{equation}
%    M^{(1)}_{gs}(k_1;R_{gs},\Omega,\omega)=-\mathrm{i}E_{\Omega} \braket{k'_1}{z|i}
%\end{equation}
%\begin{equation}
%    M^{(2)}_{bs}(k_2;R_{bs},\Omega,\omega)=-\mathrm{i}E_{\Omega}E_{\omega} \mathrm{lim}_{}\int d^3 k'_{1} \frac{\braket{k'_2}{z|k'_1}\braket{k'_1}{z|i}}{(e_1-e_1'-\mathrm{i}\epsilon)}
%\end{equation}
\begin{align}
    M^{(1)}_{g1}(k_{};R_{gs} )&=E_{\Omega_q} \sum_{\ell} \mathcal{M}^{(1)}_{\ell,0}(k_{}) Y_{\ell,0}\quad \textrm{with}
    &\mathrm{arg}[{M}_{g1}^{(1)}  ]&=\phi_q^{XUV}-\frac{\pi}{2}-\frac{\pi \ell}{2}, \\
    M^{(2)}_{b1}(k_{};R_{bs} )&=E_{\Omega_{q-2}}E_{\omega} \sum_{\ell} \mathcal{M}^{(2)}_{\ell,0}(k_{}) Y_{\ell,0}\quad \textrm{with}
    &\mathrm{arg}[{M}_{b1}^{(2)} ]&=\phi_{q-2}^{XUV}+\omega\tau-\frac{\pi \ell}{2},\\
    M^{(2)}_{b2}(k_{};R_{bs} )&=E_{\Omega_{q}}E_{-\omega} \sum_{\ell} \mathcal{M}^{(2)}_{\ell,0}(k_{}) Y_{\ell,0}\quad \textrm{with}
    &\mathrm{arg}[{M}_{b2}^{(2)} ]&=\phi_{q}^{XUV}-\omega\tau-\frac{\pi \ell}{2}.
\end{align}
Here, the atomic phase contribution including the Wigner and continuum-continuum couplings  are neglected, since they are small compared to the atto-chirp \cite{Paul1689}.
The final photoelectron momentum $k$ and angular momentum $\ell$ are the same for all paths.
$E_{\Omega/\omega}$ are the complex electric fields of the XUV and the IR pulses, $Y_{\ell,m}$ are the angular momentum and magnetic-quantum-number dependent spherical harmonics.
$\mathcal{M}^{(N)}(k)$ is a complex expression that contains the radial part of the transition-matrix element.
We can safely assume that the IR pulse $\arg[E_{\omega}]=\omega_{}(t+\tau)+IR_{chirp}$ is not chirped and we set $t=0$: $\arg[E_{\omega}]=\omega_{}\tau=\phi(\tau)$. 
%\begin{equation}
%    \mathrm{arg}[{M}_{g1}^{(1)} (k_{1})]=-\frac{\pi}{2}-\frac{\pi \ell_1}{2}+\varphi^{(a)}_{g1} +\phi_q
%\end{equation}
%\begin{equation}
%     \mathrm{arg}[{M}_{b1}^{(2)}(k_{2})]=-\frac{\pi \ell_2}{2}+\varphi^{(a)}_{b1}+\phi_{q-2}+\omega\tau
%\end{equation}
%\begin{equation}
%     \mathrm{arg}[{M}_{b2}^{(2)}(k_{3})]=-\frac{\pi \ell_3}{2}+\varphi^{(a)}_{g2}+\phi_{q}-\omega\tau
%\end{equation}
%\begin{equation}
%    \mathrm{arg}[{M}_{gs}]-\mathrm{arg}[{M}_{bs}]=-\frac{\pi}{2}-\frac{\pi \ell_1}{2}+\eta_q+ \phi_{q}+\frac{\pi \ell_1}{2}-\eta_{q-2}-\phi_{2,1}^{cc}-\phi_{q-2}-\omega\tau
%\end{equation}
%\textcolor{red}{How to deal with the spherical harmonic?}
With this, we write the phases of the   complex amplitudes of the three quantum paths for odd bands as follows: 

\begin{align}
\mathrm{arg}[g_1]&=\Theta_{gs}-\frac{\pi}{2}-\frac{\pi \ell}{2} +\phi_{q}^{XUV} &&=\Phi_{g1},\\
\mathrm{arg}[b_1]&=\Theta_{bs}-\frac{\pi \ell}{2} +\phi_{q-2}^{XUV}+2\phi(\tau) &&=\Phi_{b1}+2\phi(\tau),\\
\mathrm{arg}[b_2]&=\Theta_{bs}-\frac{\pi \ell}{2} +\phi_{q}^{XUV} &&=\Phi_{b2}.
\end{align}

\fbox{\parbox{0.9\linewidth}{To obtain Eq.\textcolor{black}{4 and 5} in the main text, we replace the complex coefficients in Eq. \ref{cohsup2} with the defined phases above and their corresponding magnitudes and get:
\begin{equation}
     \ket{\psi}= |g_1|\mbox{e}^{i\Phi_{g1}} \ket{+,-} + (|b_1|\mbox{e}^{i(\Phi_{b1}+2\phi(\tau))} +|b_2|\mbox{e}^{i\Phi_{b2}})\ket{-,+}.
\end{equation}
By setting $|g_1|\mbox{e}^{i\Phi_{g1}}= \alpha_o$, $|b_1|\mbox{e}^{i\Phi_{b1}}= \beta_o$,   and neglecting    pathway $b_2$, we get 
\begin{equation}
      \ket{\psi} = \alpha_o \ket{+,-} + \beta_o\mbox{e}^{i2\phi(\tau)} \ket{-,+}.
\end{equation}

We discuss this simplification in order to focus on the time-dependence asymmetry. However, path $b_2$ is needed in order to explain all observed structures in the experiment. 

The asymmetry parameter (Eq. S.10) in this case becomes

\begin{equation}
A=\frac{-2|\alpha_o| |\beta_o| \text{cos}\big(\text{arg}[\alpha_o]-\text{arg}[\beta_o]-2\phi(\tau)\big)}{|\alpha_o|^2+|\beta_o|^2}.  
\end{equation}
 
}}

In order to obtain an analytical expression for the asymmetry (Eq. \ref{eq:oddAsy3P}), we need to know the phase differences between the three paths:
\begin{equation}
    \begin{split}
        \Delta\phi_{g_1,b_1}&=\Delta\Theta -\frac{\pi}{2}+\Delta\phi_{q,q-2}^{XUV} -2\phi(\tau),\\
        \Delta\phi_{g_1,b_2}&=\Delta\Theta-\frac{\pi}{2},\\
        \Delta\phi_{b_1,b_2}&=+\Delta\phi_{q-2,q}^{XUV} +2\phi(\tau),
    \end{split}
\end{equation}
where $\Delta \phi_{i,j}^{XUV}=\phi_{i}^{XUV}-\phi_{j}^{XUV}$ is the chirp of the XUV pulse, and $\Delta\Theta=\Theta_{gs}-\Theta_{bs}$ is the phase difference resulting from the nuclear part (WKP phase) of the wave function.
The time dependent asymmetry parameter then reads:
\begin{equation}
    A_{odd}(\tau)=\frac{-2|g^{}_{1}||b^{}_{1}|\sin(\Delta\Theta +\Delta\phi_{q,q-2}^{XUV} -2\phi(\tau))-2|g^{}_{1}||b^{}_{2}|\sin(\Delta\Theta)}{|g_1|^2+|b_{1}|^2+|b_{2}|^2 +2 |b^{}_{1}||b^{}_{2}|\cos(\Delta\phi_{q,q-2}^{XUV} -\phi(\tau))},\label{eq:3PAsymAll}
\end{equation}
and the time-averaged asymmetry parameter is not zero: $\langle A\rangle_{\tau}=\int d\tau A(\tau)\neq 0$.

%This time-independent asymmetry is observed in the experiment \textcolor{black}{(see Fig. 2 E in the main text).}

It is worth discussing   the role of the different considered quantum paths $g_1,b_1$,and $b_2$.
If $|g_1|$ is zero, the asymmetry parameter is zero and the time-dependent population probability shows a similar form of a RABBITT-like experiment \cite{ Muller2002,Paul1689}.
If $|b_1|$ is zero, the asymmetry parameter is not time dependent.
This shows  that, in order to see a time dependence in the experiment, two interfering quantum paths need to involve two neighboring XUV frequencies.
If $|b_2|$ is zero, the asymmetry parameter is time dependent but the time average asymmetry parameter is zero. 
Only if all three paths are present, we obtain what the experiment shows:   
\begin{itemize}
    %\item a $\tau$-dependent dissociation probability $P_{odd}(\tau)$ 
    \item a $\tau$-dependent asymmetry parameter $A(\tau)$ 
    \item and a non zero time average $\langle A \rangle=\int d\tau A(\tau)\neq 0$.
\end{itemize} 
Therefore, at least these three lowest-order perturbation paths must be considered in order to model the experimental results.
However, for the discussion in the main text regarding the origin of the time-dependent asymmetry, two paths are sufficient.

\subsection{Experiment}

\subsubsection{Time-integrated OB asymmetry parameter and chirp of the XUV pulses}
\begin{figure}
\centering 
\includegraphics[keepaspectratio=true,scale=0.6]{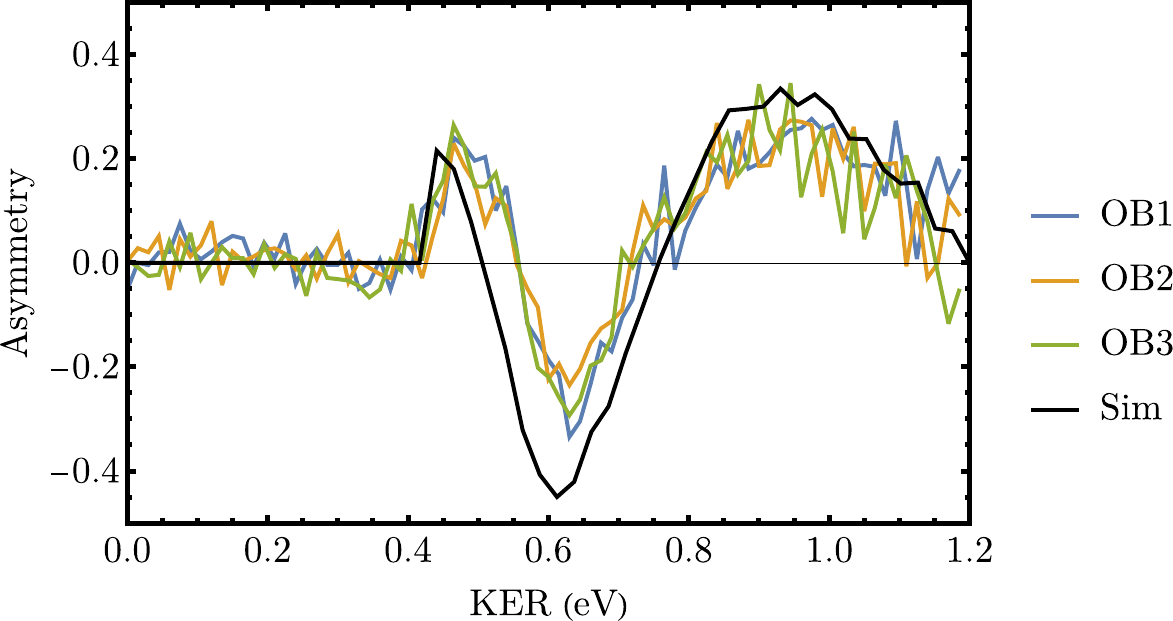}
\caption  {Experimental time-integrated odd-band asymmetry parameter $\langle A \rangle$ as a function of KER. 
Black curve shows the asymmetry parameter $A$ based on Eq. \ref{eq:3PAsymAll} with $|g_1|=\alpha_o$ and $|b1+b2|= \beta_o$.
  \label{asym}}
\end{figure}

\begin{figure}
\centering 
\includegraphics[keepaspectratio=true,scale=0.7]{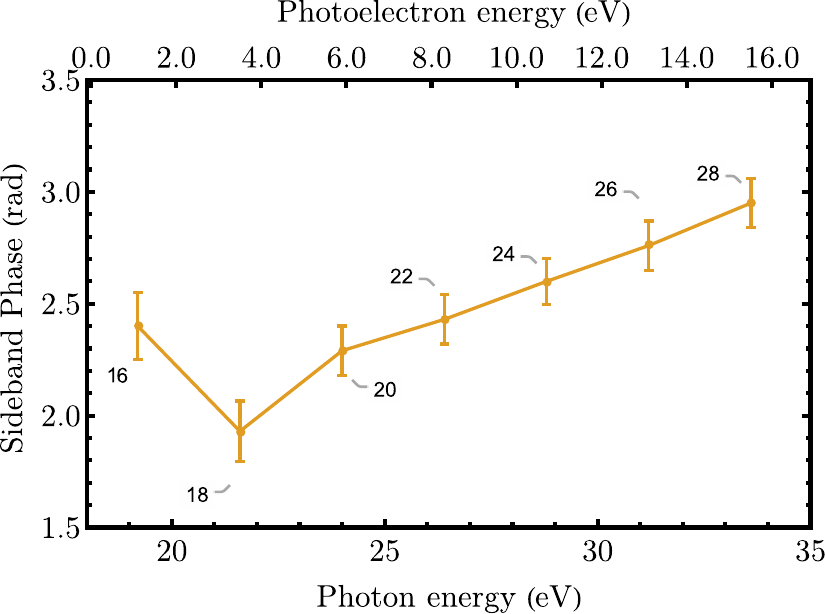}
\caption{The chirp of the XUV pulses. Shown are the relative phases of the side-bands in photoionisation of argon using a combination of XUV and IR pulses. The figure is adapted  from \cite{shobeiry_2021}.}
\label{XUVchirp}
\end{figure}

Fig. S\ref{asym} shows the experimental time-averaged asymmetry parameter $\langle A \rangle$ as a function of KER for the first three odd bands (OBs).
They correspond to harmonic 17th  (OB1), 19th (OB2), and 21st (OB3).
The black line is the theoretical curve based on the model explained above.  
{For the simulation we used the retrieved amplitudes for $g_1$, $b_1$, and $b_2$ explained in the following section 3.5. 
For $\phi(\tau)=\omega\tau$, we know the photon energy (1.2 eV) and therefore $\omega$.}
Before plotting the experimental asymmetry parameter as a function of the time delay, we add all OBs together in order to improve the statistics. 
The bands are slightly shifted in time with respect to each other mainly due to the chirp of the XUV pulse. 
This chirp is a function of the photon energy.
In order to obtain the XUV chrip, we perform a reference measurement on argon prior to the measurement on H$_2$. 
We use the photoelectron spectrum as a function of the time delay between XUV and IR pulses (also known as the RABBIT spectrum) to retrieve the chirp of the XUV pulse. Fig. S\ref{XUVchirp} shows the relative phase of the different side-bands in the RABBIT spectrum. 

\begin{figure}
\centering 
\includegraphics[keepaspectratio=true,scale=0.45]{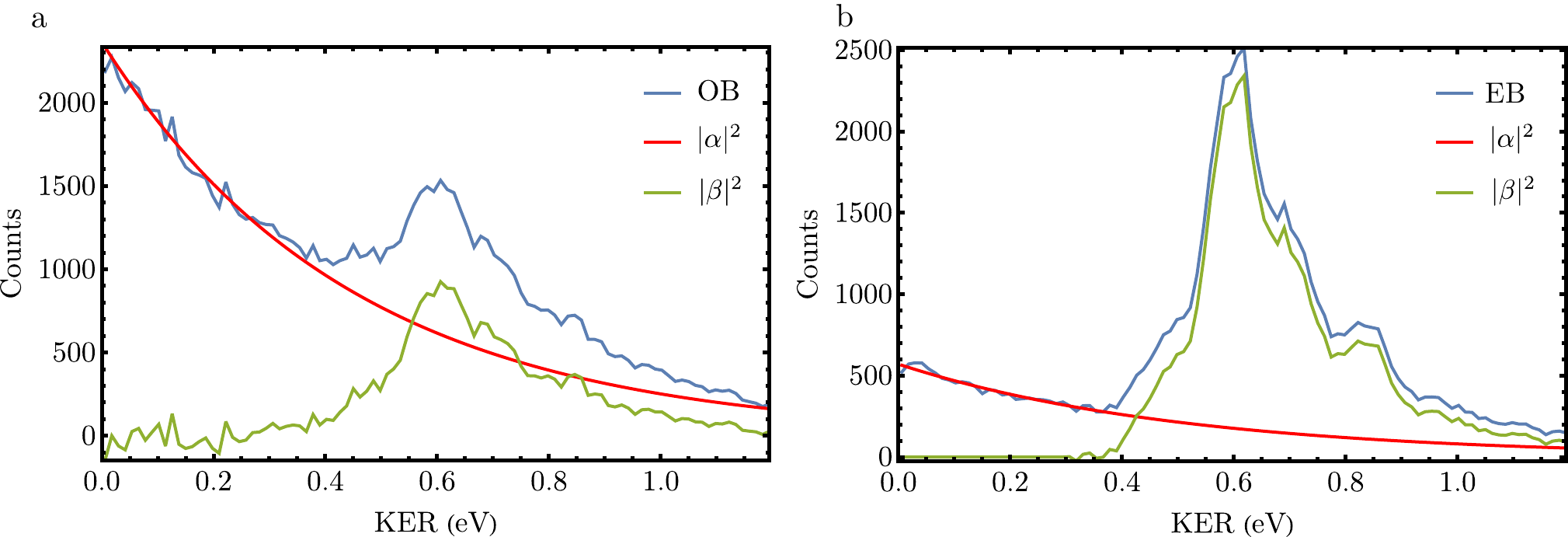}
\caption {Extraction of $\alpha$ and $\beta$ from the experimental data. a: The blue curve represents  the KER distribution of the third OB (marked with a solid line in Fig. S\ref{viblevel}-E). {b}: The blue curve represents  the KER distribution of the third EB (marked with a dashed line in Fig. S\ref{viblevel}-E). The red curve is and an exponential function fitted to the blue curve up to a KER of 0.35 eV giving us $|\alpha|^2$. The green curve is obtained by subtracting the red curve from the blue one to get $|\beta|^2$.\label{alphabeta}}
\end{figure}

%\subsubsection{Coherence of the molecular target}
%We perform experiments on a cooled gas jet of H$_2$. As a result, all H$_2$ molecules are in   a well defined ground state denoted $\ket{\color{red}+,+}$ in Fig. 1 of the main text. The H$_2$ molecules are also isolated from interacting with other molecules and atoms due to the low density of the gas target ( <$10^9$  Molecules/m$^3$). The molecules interact with a coherent laser field. The exact number of absorbed photons are known during the experiment. As a consequence, H$_2$ molecules remain coherent from the very beginning until the fragments hit the detectors. We can conclude that the final states are pure states since there is no source of incoherence during the experiment. The fact that the amplitude of the asymmetry parameter A spans at most from -0.4 to 0.4 is due to the fact that $\alpha$ and $\beta$ are not equal. In other words, the contribution of GS and BS are not the same.
\subsubsection{Coefficients $\alpha$ and $\beta$ for even and odd bands}
{
 We can retrieve the coefficients $\alpha_{o/e}$ and $\beta_{o/e}$ in Eq. 4 of the main text using the KER distribution. 
 An example is shown with the blue curve in Fig. S\ref{alphabeta} for an OB (a) as well as an EB (b). 
For the contribution of the {ground state} dissociation ($\alpha$), we fit  an exponential function of the form $f(x)=U e^{-|u|x}$  to a KER region from 0 to 0.35 eV and  obtain the exponential KER trend for the rest of the KER region up to 1.2 eV shown with the red curve in Fig S \ref{alphabeta}. 
The KER region up 0.35 eV contains only the contribution form GS dissociation. 
Then we set $|\alpha|^2=f(x)$. 
The fit is then subtracted from the experimental data to obtain the contribution of the bond softening ($\beta$). 
In case of odd bands, according to Eq. \ref{cohsup2}, we have $\beta_o=b_1+b_2$ and we set $b_1=b_2$. 
For EBs, according to Eq. \ref{cohsup3}, $\alpha_e=g_1+g_2$ and we set $g_1=g_2$.  
}

\subsubsection{Time-dependent asymmetry parameter $A$ for odd bands: Experiment and simulation}

\begin{figure*}
\centering 
\includegraphics[keepaspectratio=true,scale=0.5]{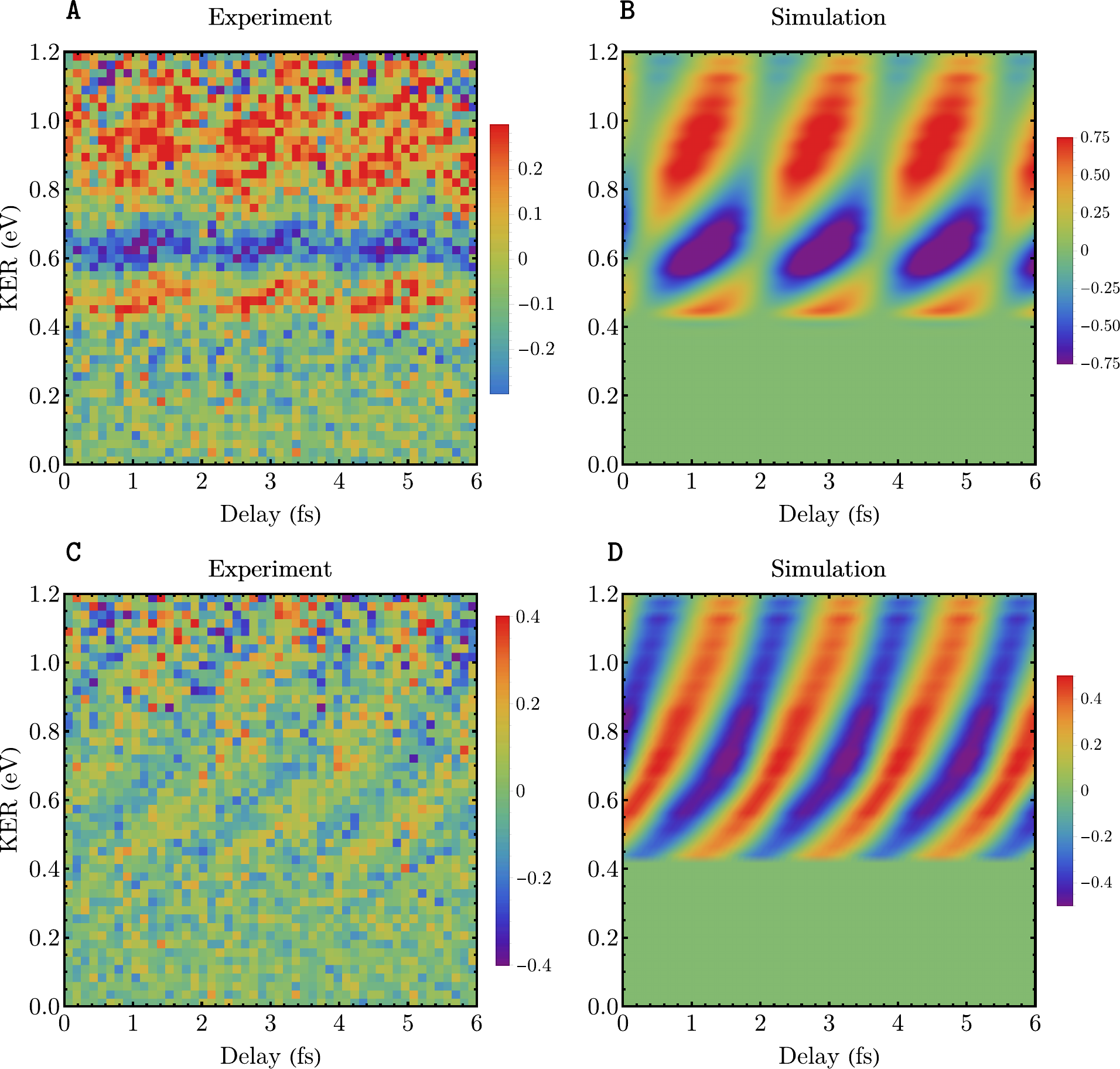}
\caption{\textbf{Time-dependent asymmetry parameter $A$ for odd bands}: A and C Experiment, B and D simulation based on the model.
The model includes the discussed main three quantum paths. 
The values of alpha and beta are obtained using the fit to the experimental data.
A and B: asymmetry parameter as a function of the time-delay between the XUV and IR pulses. 
C and D: asymmetry parameter minus the mean time-independent part ($A- \langle A \rangle$). Experiment and simulation are in fairly close agreement.
\label{Fig:oddExp}}
\end{figure*}

The final plots for the asymmetry parameter $A$ for odd bands are shown in Fig. S\ref{Fig:oddExp}-A.
Fig. \ref{Fig:oddExp}-B shows the corresponding simulated case based on the WKB approximation, the model above and the retrieved $\alpha$ and $\beta$ parameters.
Fig. S\ref{Fig:oddExp}-C and -D shows the asymmetry parameter from -A and -B minus the mean time-averaged contribution $A(\tau)- \langle A (\tau) \rangle $ for both the experiment and theory. 
The experimental and simulated results show a good agreement.

\subsection{Even-bands}
In this final section, the same procedure as for odd bands is now repeated for even bands (EBs), where the total number photons absorbed in each pathway is even. 
\begin{figure}
\centering 
\includegraphics[keepaspectratio=true,scale=0.58]{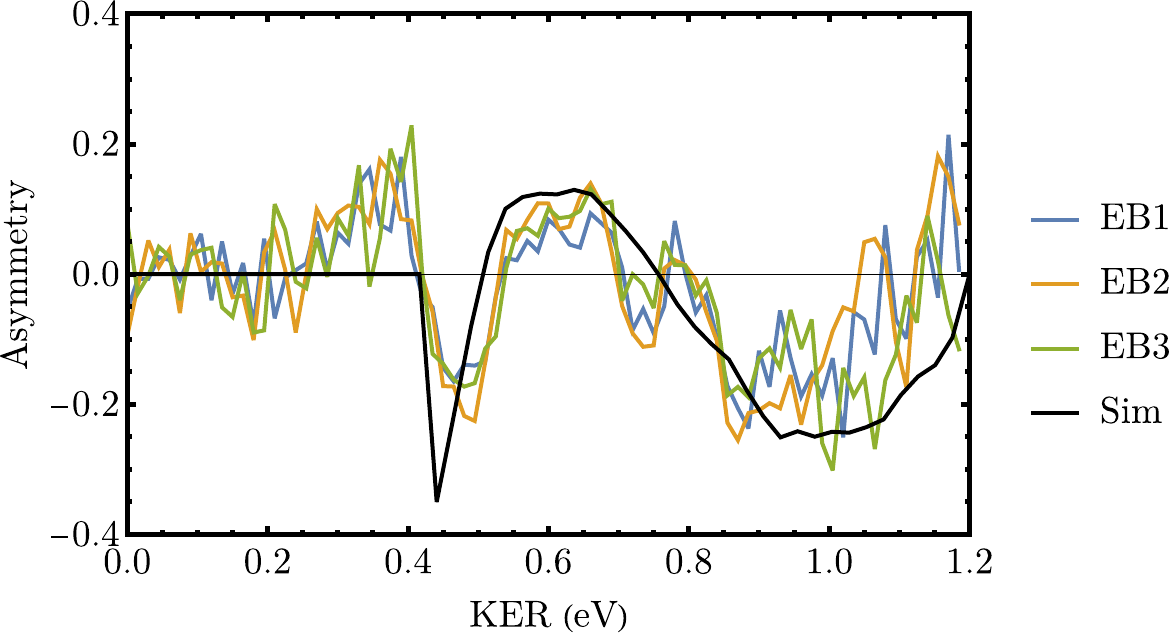}
\caption {Experimental time-integrated even-band asymmetry parameter as a function of  KER. 
The black curve shows the asymmetry $A$ based on Eq. \ref{eqEB}  with $|g_1+g_2|=\alpha_e$ and $|b1|= \beta_e$.
  \label{asymEB}}
\end{figure}
According to the pathways for EBs shown in Fig. S\ref{SF:allQP}, Eq. \ref{cohsup1} becomes:
\begin{equation}
\ket{\phi  }=  (g_1+g_2) \ket{+,+}  +b_1 \ket{-,-}.\label{cohsup3}
\end{equation} 
The population probability reads
\begin{equation}
P_{even}= |g_1+g_2+b_1|^2,
\end{equation}
and the asymmetry parameter becomes:
\begin{equation}
 A_{even}(\tau)=\frac{-2(|g^{}_{2}||b^{}_{1}|\cos(\Delta\phi_{g2,b1})+|g^{}_{1}||b^{}_{1}|\cos(\Delta\phi_{g1,b1}))}{|g_1|^2+|g_{2}|^2+|b_{1}|^2 +2 |g^{}_{1}||g^{}_{2}|\cos(\Delta\phi_{b1,b2})},\label{eqEB}
\end{equation}
with complex amplitudes: 
\begin{eqnarray}
 g_1&= &  FC_{gs}\mathrm{e}^{\mathrm{i}\theta_{gs}} \times M^{(2)}_{gs},
\end{eqnarray}
\begin{eqnarray}
 g_2&= &  FC_{gs}\mathrm{e}^{\mathrm{i}\theta_{gs}} \times M^{(2)}_{gs},
\end{eqnarray}
\begin{equation}
 b_1=   FC_{bs}\mathrm{e}^{\mathrm{i}\theta_{bs}}\times\textcolor{black}{\mathrm{e}^{\mathrm{i}\phi(\tau)}} \times M^{(1)}_{bs}.
\end{equation}
The phase difference between the nuclear parts is the same as for the odd bands, $\Delta\Theta_{even}=\Delta\Theta_{odd}$.
However, the photoelectron absorbs a different number of IR photons. The phases of the complex amplitudes are:
\begin{align}
\mathrm{arg}[g1]&=\Theta_{gs}-\frac{\pi \ell}{2}+\phi_{q-2}^{XUV}+\phi(\tau)&=\Phi_{g1}+\phi(\tau),
    \\
     \mathrm{arg}[g2]&=\Theta_{gs}-\frac{\pi \ell}{2}+\phi_{q}^{XUV}-\phi(\tau) &=\Phi_{g2}-\phi(\tau),\\
     \mathrm{arg}[b1]&=\Theta_{bs}-\frac{\pi}{2}-\frac{\pi \ell}{2}+\phi_{q-2}^{XUV}+\phi(\tau)&=\Phi_{b2}+\phi(\tau).
\end{align}
With this, the time dependent asymmetry reads:
\begin{equation}
    A_{even}(\tau)=\frac{2|g^{}_{2}||b^{}_{1}|\sin(\Delta\Theta +\Delta\phi_{q,q-2}^{XUV} -2\phi(\tau))+2|g^{}_{1}||b^{}_{1}|\sin(\Delta\Theta)}{|g_1|^2+|g_{2}|^2+|b_{1}|^2 +2 |g^{}_{1}||g^{}_{2}|\cos(\Delta\phi_{q-2,q}^{XUV} +2\phi(\tau))},\label{Eq:AvAsym_odd}
\end{equation}
and the time-averaged asymmetry parameter is not zero: $\langle A \rangle =\int d\tau A(\tau)\neq 0$.

{
Fig. \ref{asymEB}   shows the experimental time-averaged asymmetry parameter $\langle A \rangle_{\tau}$ as a function of KER for  three even bands (EBs).
They correspond to sidebands 18th  (EB1), 20th (EB2), and 22st (EB3).
The black line is the theoretical curve (Eq. \ref{Eq:AvAsym_odd}) based on the model explained above. 
}

\begin{figure*}
\centering 
\includegraphics[keepaspectratio=true,scale=0.5]{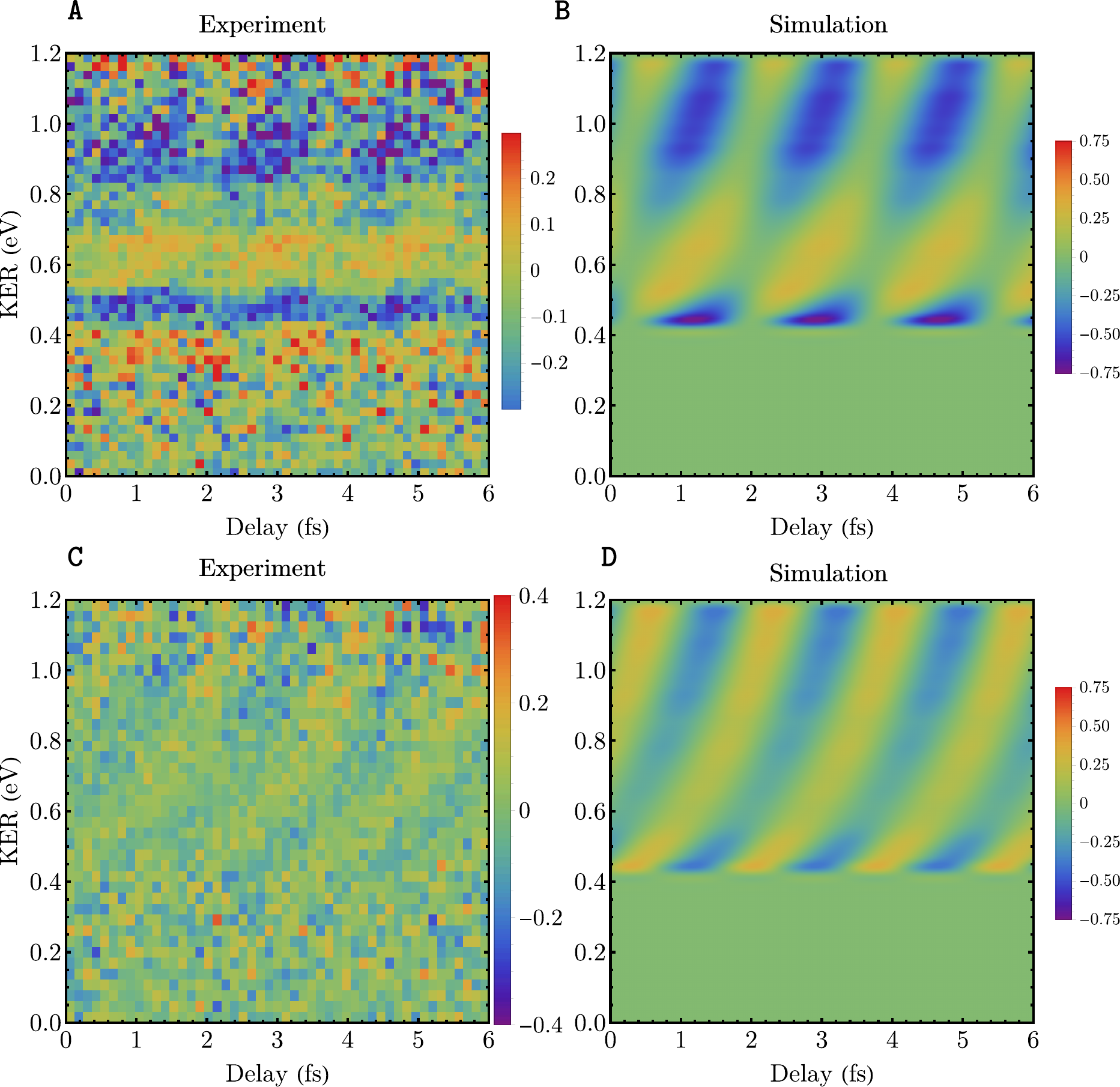}
\caption{\textbf{Time-dependent asymmetry parameter $A$ for even bands}: A and C Experiment, B and D simulation based on the model.
The model includes the discussed main three quantum paths. 
The values of alpha and beta are obtained using the fit to the experimental data.
A and B: asymmetry parameter as a function of the time-delay between the XUV and IR pulses. 
C and D: asymmetry parameter minus the mean time-independent part ($A- \langle A \rangle$). Experiment and simulation are in fairly close agreement.}\label{asymtimedepThExEBs}
\end{figure*}

The final plots for the asymmetry parameter $A$ for even bands are shown in Fig. S\ref{asymtimedepThExEBs}-A.
Fig. \ref{asymtimedepThExEBs}-B shows the corresponding simulated case based on the WKB approximation, the model above and the retrieved $\alpha$ and $\beta$ parameters.
Fig. S\ref{asymtimedepThExEBs}-C and -D shows the asymmetry parameter from -A and -B minus the mean time-averaged contribution $A(\tau)- \langle A (\tau) \rangle $ for both the experiment and theory. 
Experiment and simulation are in fairly close agreement.

%\fbox{\parbox{0.9\linewidth}
%{The Eq. 2 in the main text is obtained in a similar manner  to OBs by writing the final wave function  using the defined phases above and their corresponding amplitudes
%\begin{equation}
%    \ket{\phi}= (|g_1|\mbox{e}^{i(\Phi_{g1}+\phi(\tau))}+ |g_2|\mbox{e}^{i(\Phi_{g2}-\phi(\tau))}) \ket{+,+} + |b_1|\mbox{e}^{i(\Phi_{b1}+\phi(\tau))} \ket{-,-}.
%\end{equation}
%We neglect pathway $g_1$ for the same reasons as explained for the odd bands and get:
%\begin{equation}
%    \ket{\phi}=  |g_2|\mbox{e}^{i(\Phi_{g2}-\phi(\tau))} \ket{+,+} + |b_1|\mbox{e}^{i(\Phi_{b1}+\phi(\tau)}) \ket{-,-}. 
%\end{equation}
%By setting $|g_2|\mbox{e}^{i\Phi_{g2}}= \alpha_e$, $|b_1|\mbox{e}^{i\Phi_{b1}}= \beta_e$,   we get 
% \begin{equation}
%      \ket{\phi}= \alpha_e \ket{+,+} + \beta_e\mbox{e}^{i2\phi(\tau)} \ket{-,-}.
%\end{equation}

%}
%}

 \newpage
 \bibliographystyle{unsrt}
  \small \bibliography{ref_H2_v2,ref}

\begin{thebibliography}{10}

\bibitem{Bello2021}
R.~Y. Bello, F.~Martín, and A.~Palacios.
\newblock Attosecond laser control of photoelectron angular distributions in xuv-induced ionization of h2.
\newblock {\em Faraday Discuss.}, 228:378--393, 2021.

\bibitem{Vatasescu2023}
M.~Vatasescu.
\newblock Entanglement between electronic and vibrational degrees of freedom in a laser-driven molecular system.
\newblock {\em Phys. Rev. A}, 88:063415, 2013.

\bibitem{Ruberti2019}
M.~Ruberti.
\newblock Onset of ionic coherence and ultrafast charge dynamics in attosecond molecular ionisation.
\newblock {\em Phys. Chem. Chem. Phys.}, 21:17584--17604, 2019.

\bibitem{Nishi2019}
T.~Nishi, E.~L\"otstedt, and K.~Yamanouchi.
\newblock Entanglement and coherence in photoionization of ${\mathrm{h}}_{2}$ by an ultrashort xuv laser pulse.
\newblock {\em Phys. Rev. A}, 100:013421, 2019.

\bibitem{Carlstrom2018}
S.~Carlström, J.~Mauritsson, K.~J. Schafer, A.~L’Huillier, and M.~Gisselbrecht.
\newblock Quantum coherence in photo-ionisation with tailored xuv pulses.
\newblock {\em Journal of Physics B: Atomic, Molecular and Optical Physics}, 51(1):015201, 2017.

\bibitem{Nabekawa2023}
Y.~Nabekawa and K.~Midorikawa.
\newblock Analysis of attosecond entanglement and coherence using feasible formulae.
\newblock {\em Phys. Rev. Res.}, 5:033083, 2023.

\bibitem{Vrakking2021}
M.~J.~J. Vrakking.
\newblock Control of attosecond entanglement and coherence.
\newblock {\em Phys. Rev. Lett.}, 126:113203, 2021.

\bibitem{Kroll2022}
L.~Koll, L.~Maikowski, L.~Drescher, T.~Witting, and M.~J.~J. Vrakking.
\newblock Experimental control of quantum-mechanical entanglement in an attosecond pump-probe experiment.
\newblock {\em Phys. Rev. Lett.}, 128:043201, 2022.

\bibitem{PhysRevLett.103.213003}
M.~Kremer, B.~Fischer, B.~Feuerstein, V.~L.~B. de~Jesus, V.~Sharma, C.~Hofrichter, A.~Rudenko, U.~Thumm, C.~D. Schr\"oter, R.~Moshammer, and J.~Ullrich.
\newblock Electron localization in molecular fragmentation of h2 by carrier-envelope phase stabilized laser pulses.
\newblock {\em Phys. Rev. Lett.}, 103:213003, 2009.

\bibitem{Sansone2010}
G.~Sansone, F.~Kelkensberg, J.~F. P{\'e}rez-Torres, F.~Morales, M.~F. Kling, W.~Siu, O.~Ghafur, P.~Johnsson, M.~Swoboda, E.~Benedetti, F.~Ferrari, F.~L{\'e}pine, J.~L. Sanz-Vicario, S.~Zherebtsov, I.~Znakovskaya, A.~L'Huillier, M.~Yu. Ivanov, M.~Nisoli, F.~Mart{\'i}n, and M.~J.~J. Vrakking.
\newblock Electron localization following attosecond molecular photoionization.
\newblock {\em Nature}, 465(7299):763--766, 2010.

\bibitem{Kling246}
M.~F. Kling, Ch. Siedschlag, A.~J. Verhoef, J.~I. Khan, M.~Schultze, Th. Uphues, Y.~Ni, M.~Uiberacker, M.~Drescher, F.~Krausz, and M.~J.~J. Vrakking.
\newblock Control of electron localization in molecular dissociation.
\newblock {\em Science}, 312(5771):246--248, 2006.

\bibitem{PhysRevLett.104.023001}
K.~P. Singh, F.~He, P.~Ranitovic, W.~Cao, S.~De, D.~Ray, S.~Chen, U.~Thumm, A.~Becker, M.~M. Murnane, H.~C. Kapteyn, I.~V. Litvinyuk, and C.~L. Cocke.
\newblock Control of electron localization in deuterium molecular ions using an attosecond pulse train and a many-cycle infrared pulse.
\newblock {\em Phys. Rev. Lett.}, 104:023001, 2010.

\bibitem{Fischer2013}
A.~Fischer, A.~Sperl, P.~C\"orlin, M.~Sch\"onwald, H.~Rietz, A.~Palacios, A.~Gonz\'alez-Castrillo, F.~Martín, T.~Pfeifer, J.~Ullrich, A.~Senftleben, and R.~Moshammer.
\newblock Electron localization involving doubly excited states in broadband extreme ultraviolet ionization of h2.
\newblock {\em Phys. Rev. Lett.}, 110:213002, 2013.

\bibitem{Vrakking2022}
M.~J.~J. Vrakking.
\newblock Ion-photoelectron entanglement in photoionization with chirped laser pulses.
\newblock {\em Journal of Physics B: Atomic, Molecular and Optical Physics}, 55(13):134001, 2022.

\bibitem{Martin629}
F.~Martín, J.~Fernandez, T.~Havermeier, L.~Foucar, Th. Weber, K.~Kreidi, M.~Schoeffler, L.~Schmidt, T.~Jahnke, O.~Jagutzki, A.~Czasch, E.~P. Benis, T.~Osipov, A.~L. Landers, A.~Belkacem, M.~H. Prior, H.~Schmidt-B{\"o}cking, C.~L. Cocke, and R.~Doerner.
\newblock Single photon-induced symmetry breaking of h2 dissociation.
\newblock {\em Science}, 315(5812):629--633, 2007.

\bibitem{remi}
J.~Ullrich, R.~Moshammer, A.~Dorn, R.~Dörner, L.~Ph.~H. Schmidt, and H.~Schmidt-Böcking.
\newblock Recoil-ion and electron momentum spectroscopy reaction-microscopes.
\newblock {\em Reports on Progress in Physics}, 66(9):1463, 2003.

\bibitem{Paul1689}
P.~M. Paul, E.~S. Toma, P.~Breger, G.~Mullot, F.~Aug{\'e}, Ph. Balcou, H.~G. Muller, and P.~Agostini.
\newblock Observation of a train of attosecond pulses from high harmonic generation.
\newblock {\em Science}, 292(5522):1689--1692, 2001.

\bibitem{Hem2023}
H.~Srinivas, F.~Shobeiry, D.~Bharti, T.~Pfeifer, R.~Moshammer, , and A.~Harth.
\newblock High-repetition rate attosecond beamline for multi-particle coincidence experiments.
\newblock {\em Optics Express}, 30(8):13630, 2022.

\bibitem{Cattaneo2018}
L.~Cattaneo, J.~Vos, R.~Y. Bello, A.~Palacios, S.~Heuser, L.~Pedrelli, M.~Lucchini, C.~Cirelli, F.~Mart{\'i}n, and U.~Keller.
\newblock Attosecond coupled electron and nuclear dynamics in dissociative ionization of h2.
\newblock {\em Nature Physics}, 14(7):733--738, 2018.

\bibitem{PhysRevLett.64.1883}
P.~H. Bucksbaum, A.~Zavriyev, H.~G. Muller, and D.~W. Schumacher.
\newblock Softening of the ${\mathrm{h}}_{2}^{+}$ molecular bond in intense laser fields.
\newblock {\em Phys. Rev. Lett.}, 64:1883--1886, 1990.

\bibitem{DORNER200095}
R.~Dörner, V.~Mergel, O.~Jagutzki, L.~Spielberger, J.~Ullrich, R.~Moshammer, and H.~Schmidt-Böcking.
\newblock Cold target recoil ion momentum spectroscopy: a ‘momentum microscope’ to view atomic collision dynamics.
\newblock {\em Physics Reports}, 330(2):95--192, 2000.

\bibitem{0034-4885-66-9-203}
J~Ullrich, R~Moshammer, A~Dorn, R~Dörner, L~Ph~H Schmidt, and H~Schmidt-Böcking.
\newblock Recoil-ion and electron momentum spectroscopy: reaction-microscopes.
\newblock {\em Reports on Progress in Physics}, 66(9):1463, 2003.

\bibitem{shobeiry_2021}
Farshad Shobeiry.
\newblock {\em Attosecond Electron-Nuclear Dynamics in Photodissociation of H2 and D2}.
\newblock PhD thesis, 2021.

\bibitem{zare}
Richard~N. Zare.
\newblock Photoejection dynamics,.
\newblock {\em Mol. Photochem}, 1972.

\bibitem{doi:10.1063/1.465352}
Y.~M. Chung, E.‐M. Lee, T.~Masuoka, and James A.~R. Samson.
\newblock Dissociative photoionization of h2 from 18 to 124 ev.
\newblock {\em The Journal of Chemical Physics}, 99(2):885--889, 1993.

\bibitem{Fischer_2013}
Andreas Fischer, Alexander Sperl, Philipp Cörlin, Michael Schönwald, Sebastian Meuren, Joachim Ullrich, Thomas Pfeifer, Robert Moshammer, and Arne Senftleben.
\newblock Measurement of the autoionization lifetime of the energetically lowest doubly excited state in h2using electron ejection asymmetry.
\newblock {\em Journal of Physics B: Atomic, Molecular and Optical Physics}, 47(2):021001, dec 2013.

\bibitem{doi:10.1002/cphc.201200974}
Alicia Palacios, Johannes Feist, Alberto González-Castrillo, José~Luis Sanz-Vicario, and Fernando Martín.
\newblock Autoionization of molecular hydrogen: Where do the fano lineshapes go?
\newblock {\em ChemPhysChem}, 14(7):1456--1463, 2013.

\bibitem{PhysRevA.48.2145}
Eric~E. Aubanel, Jean-Marc Gauthier, and Andr\'e~D. Bandrauk.
\newblock Molecular stabilization and angular distribution in photodissociation of h2+ in intense laser fields.
\newblock {\em Phys. Rev. A}, 48:2145--2152, Sep 1993.

\bibitem{PhysRevA.77.063403}
Manohar Awasthi, Yulian~V. Vanne, Alejandro Saenz, Alberto Castro, and Piero Decleva.
\newblock Single-active-electron approximation for describing molecules in ultrashort laser pulses and its application to molecular hydrogen.
\newblock {\em Phys. Rev. A}, 77:063403, Jun 2008.

\bibitem{PhysRevA.103.022834}
Divya Bharti, David Atri-Schuller, Gavin Menning, Kathryn~R. Hamilton, Robert Moshammer, Thomas Pfeifer, Nicolas Douguet, Klaus Bartschat, and Anne Harth.
\newblock Decomposition of the transition phase in multi-sideband schemes for reconstruction of attosecond beating by interference of two-photon transitions.
\newblock {\em Phys. Rev. A}, 103:022834, Feb 2021.

\bibitem{Muller2002}
H.G. Muller.
\newblock Reconstruction of attosecond harmonic beating by interference of two-photon transitions.
\newblock {\em Applied Physics B}, 74(1):s17--s21, 2002.

\end{thebibliography}


\begin{thebibliography}{10}

\bibitem{Bello2021AttosecondLC}
R.Y.~Bello, F.~Mart{\'i}n, and A.~Palacios.
\newblock Attosecond laser control of photoelectron angular distributions in
  xuv-induced ionization of ${\mathbf{H}}_{2}$.
\newblock {\em Faraday discussions}, 2021.

\bibitem{Martin629}
F.~Martin, J.~Fernandez, T.~Havermeier, L.~Foucar, Th.~Weber, K.~Kreidi,
  M.~Schoeffler, L.~Schmidt, T.~Jahnke, O.~Jagutzki, A.~Czasch, E.~P. Benis,
  T.~Osipov, A.~L. Landers, A.~Belkacem, M.~H. Prior, H.~Schmidt-B{\"o}cking,
  C.~L. Cocke, and R.~Doerner.
\newblock Single photon-induced symmetry breaking of ${\mathbf{H}}_{2}$ dissociation.
\newblock {\em Science}, 315(5812):629--633, 2007.

\bibitem{PhysRevLett.110.213002}
A.~Fischer, A.~Sperl, P.~C\"orlin, M.~Sch\"onwald, H.~Rietz, A.~Palacios, A.~Gonz\'alez-Castrillo, F.~Mart\'{\i}n,
  T.~Pfeifer, J.~Ullrich, A.~Senftleben, and R.~Moshammer.
\newblock Electron localization involving doubly excited states in broadband
  extreme ultraviolet ionization of ${\mathbf{H}}_{2}$.
\newblock {\em Phys. Rev. Lett.}, 110:213002, 2013.

\bibitem{PhysRevLett.103.213003}
Manuel Kremer, Bettina Fischer, Bernold Feuerstein, Vitor L.~B. de~Jesus,
  V.~Sharma, C.~Hofrichter, A.~Rudenko, U.~Thumm, C.~Schr\"oter, R.~Moshammer, and J.~Ullrich.
\newblock Electron localization in molecular fragmentation of h2 by
  carrier-envelope phase stabilized laser pulses.
\newblock {\em Phys. Rev. Lett.}, 103:213003, 2009.

\bibitem{Hem2023}
H.~Srinivas, F.~Shobeiry, D.~Bharti, T.~Pfeifer, R.~Moshammer, and A.~Harth. 
\newblock High-repetition rate attosecond beamline for multi-particle coincidence experiments
\newblock {\em Optics Express}, 30(8): 13630-13646, 2022. 

\bibitem{Sansone2010}
G.~Sansone, F.~Kelkensberg, J.~F. P{\'e}rez-Torres, F.~Morales, M.~F. Kling,
  W.~Siu, O.~Ghafur, P.~Johnsson, M.~Swoboda, E.~Benedetti, F.~Ferrari,
  F.~L{\'e}pine, J.~L. Sanz-Vicario, S.~Zherebtsov, I.~Znakovskaya,
  A.~L'Huillier, M.~Yu. Ivanov, M.~Nisoli, F.~Mart{\'i}n, and M.~J.~J.
  Vrakking.
\newblock Electron localization following attosecond molecular photoionization.
\newblock {\em Nature}, 465(7299):763--766, 2010.

\bibitem{Kling246}
M.~F. Kling, Ch. Siedschlag, A.~J. Verhoef, J.~I. Khan, M.~Schultze, Th.
  Uphues, Y.~Ni, M.~Uiberacker, M.~Drescher, F.~Krausz, and M.~J.~J. Vrakking.
\newblock Control of electron localization in molecular dissociation.
\newblock {\em Science}, 312(5771):246--248, 2006.

\bibitem{remi}
J.~Ullrich, R.~Moshammer, A.~Dorn, R.~Dörner, L.Ph.H.~Schmidt, and
  H.~Schmidt-Böcking.
\newblock Recoil-ion and electron momentum spectroscopy reaction-microscopes.
\newblock {\em Reports on Progress in Physics}, 66(9):1463, 2003.

\bibitem{Paul1689}
P.~M. Paul, E.~S. Toma, P.~Breger, G.~Mullot, F.~Aug{\'e}, Ph.~Balcou, H.~G.
  Muller, and P.~Agostini.
\newblock Observation of a train of attosecond pulses from high harmonic
  generation.
\newblock {\em Science}, 292(5522):1689--1692, 2001.

\bibitem{Cattaneo2018}
L.~Cattaneo, J.~Vos, R.~Y. Bello, A.~Palacios, S.~Heuser, L.~Pedrelli,
  M.~Lucchini, C.~Cirelli, F.~Mart{\'i}n, and U.~Keller.
\newblock Attosecond coupled electron and nuclear dynamics in dissociative
  ionization of h2.
\newblock {\em Nature Physics}, 14(7):733--738, 2018.

\bibitem{PhysRevLett.64.1883}
P.~H. Bucksbaum, A.~Zavriyev, H.G.~Muller, and D.W.~Schumacher.
\newblock Softening of the ${\mathrm{h}}_{2}^{+}$ molecular bond in intense
  laser fields.
\newblock {\em Phys. Rev. Lett.}, 64:1883--1886, 1990.

\bibitem{DORNER200095}
R.~Dörner, V.~Mergel, O.~Jagutzki, L.~Spielberger, J.~Ullrich, R.~Moshammer,
  and H.~Schmidt-Böcking.
\newblock Cold target recoil ion momentum spectroscopy: a ‘momentum
  microscope’ to view atomic collision dynamics.
\newblock {\em Physics Reports}, 330(2):95 -- 192, 2000.

\end{thebibliography}

\newpage

\end{document}